\documentclass[lettersize,journal]{IEEEtran}
\usepackage{amsmath,amsfonts}
\usepackage{algpseudocode}
\usepackage{algorithm}
\usepackage{amsmath,amsfonts}
\usepackage{algpseudocode}
\usepackage{amssymb}
\usepackage{algorithm}
\usepackage{array}
\usepackage[caption=false,font=normalsize,labelfont=sf,textfont=sf]{subfig}
\usepackage{textcomp}
\usepackage{stfloats}
\usepackage{url}
\usepackage{verbatim}
\usepackage{color,soul}
\usepackage{graphicx}
\usepackage{hyperref}
\usepackage{cite}
\usepackage{makecell}
\usepackage[nolist]{acronym}
\usepackage{tikz}
\usetikzlibrary{arrows.meta}
\usepackage{tikz}
\usepackage{graphicx}
\usetikzlibrary{calc}
\usepackage{subcaption}        
\usepackage{booktabs}  
\usepackage[normalem]{ulem}

\usepackage{xcolor,cite,etoolbox}
\makeatletter 
\pretocmd\@bibitem{\color{black}\csname keycolor#1\endcsname}{}{\fail}
\newcommand\citecolor[1]{\@namedef{keycolor#1}{\color{blue}}}
\makeatother

\begin{acronym}
\acro{HD}{half-duplex}
\acro{FD}{full-duplex}
\acro{MIMO}{multiple-input, multiple-output}
\acro{AWGN}{additive white Gaussian noise}
\acro{KPI}{key performance indicator}
\acro{KPIs}{key performance indicators}  
\acro{D2D}{device-to-device}
\acro{ISAC}{integrated sensing and communications}
\acro{DFRC}{dual-functional radar and communication}
\acro{AoA}{angle of arrival}
\acro{AoAs}{angles of arrival}
\acro{ToA}{time of arrival}
\acro{EVM}{error vector magnitude}
\acro{CPI}{coherent processing interval}
\acro{eMBB}{enhanced mobile broadband}
\acro{URLLC}{ultra-reliable low latency communications}
\acro{mMTC}{massive machine type communications}
\acro{QCQP}{quadratic constrained quadratic programming}
\acro{BnB}{branch and bound}
\acro{SER}{symbol error rate}
\acro{LFM}{linear frequency modulation}
\acro{BS}{base station}
\acro{UE}{user equipment}
\acro{DL}{downlink}
\acro{UL}{uplink}
\acro{CS}{compressed sensing}
\acro{PR}{passive radar}
\acro{LMS}{least mean squares}
\acro{NLMS}{normalized least mean squares}
\acro{RIS}{reconfigurable intelligent surface}
\acro{RISs}{reconfigurable intelligent surfaces}
\acro{IRS}{intelligent reflecting surface}
\acro{LISA}{large intelligent surface/antennas}
\acro{OFDM}{orthogonal frequency-division multiplexing}
\acro{EM}{electromagnetic}
\acro{mmWave}{milli-meter wave}
\acro{MSE}{mean squared error}
\acro{XR}{extended reality}
\acro{SNR}{signal-to-noise ratio}
\acro{SRP}{successful recovery probability}
\acro{CDF}{cumulative distribution function}
\acro{ULA}{uniform linear array}
\acro{RSS}{received signal strength}
\acro{FIM}{Fisher information matrix}
\acro{BB}{Barankin bound}
\acro{SNDR}{signal-to-noise-and-distortion ratio}
\acro{DPI}{direct path interference}
\acro{RPI}{reflected path interference}
\acro{CRB}{Cram\'er-Rao bound}
\acro{ML}{maximum likelihood}
\acro{DE}{direct echo}
\acro{RE}{RIS-resulted echo}
\acro{PI}{path interference}
\acro{ADC}{analog-to-digital converter}
\acro{RF}{radio frequency}
\acro{BCD}{block coordinate descent}
\acro{BCCD}{block cyclic coordinate descent}
\acro{RCG}{Riemannian conjugate gradient}
\acro{SDP}{semi-definite program}
\acro{EVD}{eigenvalue decomposition}
\acro{RCS}{radar cross-section}
\acro{STAR}{simultaneous transmitting and reflecting}
\acro{LNA}{low noise amplifier}
\acro{NOMA}{non-orthogonal multiple access}
\acro{FM}{frequency modulation}
\acro{SINR}{signal to interference plus noise ratio}
\acro{V2X}{Vehicle-to-Everything}
\acro{3GPP}{3rd Generation Partnership Project}
\acro{AMC}{Adaptive Modulation and Coding}
\acro{AWGN}{Additive White Gaussian Noise}
\acro{AOD}{Angle of Departure}
\acro{AOA}{Angle of Arrival}
\acro{BS}{Base Station}
\acro{BER}{Bit Error Rate}
\acro{CIR}{Channel Impulse Response}
\acro{CS}{Channel Simulator}
\acro{CI}{Close-in}
\acro{CDL}{Cluster Delay Line}
\acro{Co-Pol}{Co-Polarization}
\acro{CIH}{Close-in PL model with Height dependence}
\acro{Cross-Pol}{Cross-Polarization}
\acro{CW}{Continuous Wave}
\acro{EDGE}{Enhanced Data rates for GSM Evolution}
\acro{ETSI}{European Telecommunications Standards Institute}
\acro{GSM}{Global System for Mobile Communications}
\acro{GUI}{Graphical User Interface}
\acro{HPBW}{Half Power Beamwidth}
\acro{HSPA}{High Speed Packet Access}
\acro{HST}{High Speed Train}
\acro{H-H}{Horizontal–Horizontal}
\acro{HBF}{Hybrid Beamforming}
\acro{InH}{Indoor Hotspot}
\acro{InF}{Indoor Factory}
\acro{ITU}{International Telecommunication Union}
\acro{LSP}{Large-Scale Parameter}
\acro{LOS}{Line of Sight}
\acro{LLS}{Link Level Simulator}
\acro{LTE}{Long-Term Evolution}
\acro{MU}{Multi-user}
\acro{NS}{Network Simulator}
\acro{NR}{New Radio}
\acro{NLOS}{Non Line of Sight}
\acro{NTN}{Non Terrestrial Networks}
\acro{NOMA}{Non-Orthogonal Multiple Access}
\acro{NYUSIM}{NYU Channel Model Simulator}
\acro{O2I}{Outdoor-to-Indoor}
\acro{PL}{Path Loss}
\acro{PLE}{Path Loss Exponent}
\acro{RF}{Radio Frequency}
\acro{Rx}{Receiver}
\acro{RMS}{Root Mean Square}
\acro{RMa}{Rural Macrocell}
\acro{SL}{Spatial Lobe}
\acro{SINR}{Signal-to-Interference and Noise Ratio}
\acro{SCM}{Stochastic Channel Model}
\acro{SLS}{System Level Simulator}
\acro{SE}{Spectral Efficiency}
\acro{SU}{Single-user}
\acro{SISO}{Single Input Single Output}
\acro{TC}{Time Cluster}
\acro{T-R}{Transmitter–Receiver}
\acro{Tx}{Transmitter}
\acro{FR3}{Frequency Range 3}
\acro{UAV}{Unmanned Aerial Vehicle}
\acro{UT}{User Terminal}
\acro{ULA}{Uniform Linear Array}
\acro{URA}{Uniform Rectangular Array}
\acro{UM}{Ultra Massive}
\acro{V2X}{Vehicle-to-everything}
\acro{V-V}{Vertical–Vertical}
\acro{V-H}{Vertical–Horizontal}
\acro{WCDMA}{Wideband Code Division Multiple Access}
\acro{XPD}{Cross-Polarization Discriminator}
\acro{CPD}{Co-Polarization Discrimination}
\acro{FSPL}{Free Space Path Loss}
\acro{SNR}{Signal-to-Noise Ratio}
\acro{PDP}{Power Delay Profile}
\acro{dB}{decibel}
\acro{ZoA}{zenith angle of arrival}
\acro{ZoD}{zenith angle of departure}
\acro{AoA}{angle of arrival}
\acro{AoD}{angle of departure}
\acro{ToA}{time of arrival}
\acro{AR}{augmented reality}
\acro{VR}{virtual reality}
\acro{MPC}{multipath component}
\acro{IoT}{internet-of-things}
\acro{AI}{artificial intelligence}
\acro{RF}{radiofrequency}
\acro{CSI}{channel state information}
\acro{KKT}{Karush-Kuhn-Tucker}
\acro{SAR}{Synthetic-aperture radar}
\acro{LoS}{line-of-sight}
\acro{UPA}{uniform planar array}
\acro{CD}{Chamfer distance}
\acro{LMMSE}{linear minimum mean-square error}
\acro{SSIM}{structural similarity index measure}
\acro{ERP}{equivalent reflection point}
\acro{MVF}{multi-vantage fusion}
\end{acronym}

\usetikzlibrary{positioning}
\usepackage{fancyhdr}
\fancypagestyle{firststyle}{
	\fancyhf{}
	\fancyhead[L]{A. Bazzi, M. Ying, O. Kanhere, T. S. Rappaport and M. Chafii, ``ISAC Imaging by Channel State Information using Ray Tracing for Next Generation 6G'', accepted in \textit{IEEE Journal on Selected Topics in Electromagnetics, Antennas and Propagation (JSTEAP)}, 2025. (Invited paper)}

}

\def\BibTeX{{\rm B\kern-.05em{\sc i\kern-.025em b}\kern-.08em
    T\kern-.1667em\lower.7ex\hbox{E}\kern-.125emX}}
\begin{document}
\title{ISAC Imaging by Channel State Information using Ray Tracing for Next Generation 6G}
\author{Ahmad Bazzi, Mingjun Ying, Ojas Kanhere, Theodore S. Rappaport and Marwa Chafii 
\thanks{Ahmad Bazzi and Marwa Chafii are with the Engineering Division, New York University Abu Dhabi (NYUAD), 129188, UAE
(email: \{ahmad.bazzi,marwa.chafii\}@nyu.edu)}
\thanks{Ahmad Bazzi, Minjung Ying, Theodore S. Rappaport and Marwa Chafii are with NYU WIRELESS, NYU Tandon School of Engineering, Brooklyn, 11201, NY, USA (email: \{my2770,tsr\}@nyu.edu).}
\thanks{Ojas Kanhere is with AT\&T.
(email: ok067s@att.com).}
\thanks{Manuscript received xxx}}
\markboth{Journal of \LaTeX\ Class Files,~Vol.~14, No.~8, August~2021}%
{Shell \MakeLowercase{\textit{et al.}}: A Sample Article Using IEEEtran.cls for IEEE Journals}


\maketitle
\thispagestyle{firststyle}

\begin{abstract}
\textcolor{black}{Integrated sensing and communications (ISAC) is emerging as a cornerstone technology for sixth generation (6G) wireless systems, unifying connectivity and environmental mapping through shared hardware, spectrum, and waveforms. 
The following paper presents an ISAC imaging framework utilizing channel state information (CSI) per-path components, transmitter (TX) positions, and receiver (RX) positions obtained from the calibrated NYURay ray tracer at 6.75 GHz in the upper mid-band. 
\textcolor{black}{Our work shows how each resolvable multipath component can be extracted from CSI estimation and cast into an equivalent three-dimensional reflection point by fusing its angle and delay information, which is useful and challenging for multi-bounce reflections.} The primary contribution of the paper is the two-segment reflection point optimization algorithm, which independently estimates the path lengths from the TX position and RX position to an equivalent reflection point \textcolor{black}{(ERP)} on the object surface, thus enabling precise geometric reconstruction. 
Subsequently, we aggregate the \textcolor{black}{ERPs} derived from multiple pairs of TX and RX positions, generating dense three dimensional point clouds representing the objects in the channel. 
Experimental results validate that the proposed ISAC imaging framework accurately reconstructs object surfaces, edges, and curved features.
 To the best of our knowledge, this paper provides the first demonstration of multi bounce ISAC imaging using wireless ray tracing at 6.75 GHz.}
\end{abstract}

\section{Introduction}
\label{sec:introduction}
\textcolor{black}{Sixth-generation ($6$G) networks will push radio links into multi-gigahertz bandwidths and higher carrier frequencies, in addition to sub-millisecond latency and sensing capabilities}\cite{10041914,10018908,Umer2025}.
\textcolor{black}{A key enabler is \ac{ISAC} \cite{chen2025beyond}, which aims at} unifying radar-class environmental awareness with high throughput data exchange on the same spectrum \cite{Zhang2024292}, hardware \cite{10694318}, and waveforms \cite{10061453,10904093,10549948,10771629,10926558,10707080,10891750,10681468} with little or no additional overhead. 
\textcolor{black}{In particular, ISAC architectures can increase spectral and energy efficiency \cite{10726912,10494224,10777502,10646355}, allow for more compact hardware, support novel and valuable applications for the end-user or network operators, and deliver reciprocal performance gains to both sensing and communication functions.}
\textcolor{black}{The convergence between sensing and communications underpins three flagship 6G use cases: city-scale digital twins\cite{liu2025uncovering,10489861,da20236g}, autonomous vehicles\cite{10283734,10502156,10494224}, and \ac{XR} services, all of which require high accurate 3D mapping and ultra-reliable connectivity.}
%

\begin{figure}[t]
\centering
\begin{tikzpicture}[>=Latex, thick, font=\small]

\coordinate (Tx) at (-3.2,  1.0);   
\coordinate (Rx) at ( 3.2,  1.0);   

\def\s{1.8}          
\def\shift{0.8}      

\coordinate (C) at (-\s/2,-\s/2);  

\coordinate (FLL) at (C);
\coordinate (FLR) at ($(C)+(\s,0)$);
\coordinate (FUL) at ($(C)+(0,\s)$);
\coordinate (FUR) at ($(C)+(\s,\s)$);

\coordinate (RLL) at ($(FLL)+(\shift,\shift)$);
\coordinate (RLR) at ($(FLR)+(\shift,\shift)$);
\coordinate (RUL) at ($(FUL)+(\shift,\shift)$);
\coordinate (RUR) at ($(FUR)+(\shift,\shift)$);

\fill[blue!20!white, opacity=0.25]  (RLL)--(RLR)--(RUR)--(RUL)--cycle;    
\fill[blue!15!white, opacity=0.25]  (FUL)--(FUR)--(RUR)--(RUL)--cycle;    
\fill[blue!10!white, opacity=0.25]  (FLR)--(FUR)--(RUR)--(RLR)--cycle;    

\draw[very thick]  (FLL)--(FLR)--(FUR)--(FUL)--cycle;
\draw[very thick]  (RLL)--(RLR)--(RUR)--(RUL)--cycle;
\draw[very thick]  (FLL)--(RLL) (FLR)--(RLR) (FUR)--(RUR) (FUL)--(RUL);


\draw[fill=black] (Tx) circle(2pt) node[above] {$\mathbf p_{\mathrm{Tx}_i}$};
\draw[fill=black] (Rx) circle(2pt) node[above] {$\mathbf p_{\mathrm{Rx}_i}$};

\foreach \k/\Ax/\Ay/\Bx/\By in {
      1/ -1.00/ 0.20/  1.00/ 0.75,
      2/ -0.30/-0.50/  0.90/-0.20,
      3/  0.15/ 1.10/  0.60/ 0.05
}{
  \coordinate (A\k) at (\Ax,\Ay);
  \coordinate (B\k) at (\Bx,\By);
  \coordinate (ERP\k) at ($(A\k)!0.5!(B\k)$);

  \draw[->, blue!70, line width=0.9pt] (Tx) -- (A\k);
  \draw[->, blue!70, line width=0.9pt] (B\k) -- (Rx);

  \draw[fill=black] (A\k) circle(1.8pt);
  \draw[fill=black] (B\k) circle(1.8pt);
  \draw[fill=white, draw=black, thick] (ERP\k) circle(2pt);
}

\end{tikzpicture}
\caption{RF imaging of a cubic structure, and modeling by equivalent reflection points.}
\label{fig:geom_model_cube_centered}
\end{figure}
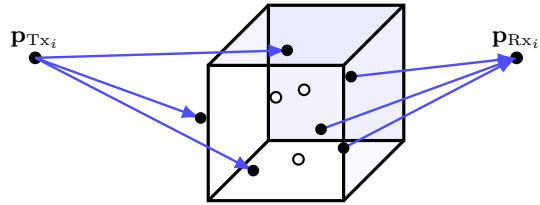

The convergence of communication and sensing capabilities in next generation wireless systems has sparked interest in using radio signals to \textit{``see''} and reconstruct complex environments for applications such as digital twins and smart infrastructures. 
\textcolor{black}{Unlike cameras, radio waves penetrate foliage, walls, and work in all lighting, where their multipath phases encode fine-scale geometry, which can further allow position and shape inference of objects without any additional hardware beyond a standard phased array.} 
Prior work has shown that \ac{CSI}-based imaging can locate human targets and identify hand gestures, but typically assumes single-bounce reflections, limiting fidelity in representing complex objects \cite{chimwere2024gesture,wang2019survey,studer2018channel,rahman2022location,tian2018performance}.

\textbf{Objective and scope.} 
The goal of this work is to use readily available, standardized \ac{CSI} in order to exploit knowledge of the multipath channel.
We turn every resolvable multipath component, readily available after standard \ac{CSI} estimation, into an \textbf{equivalent three-dimensional reflection point}. 
A base station can build a live map \cite{10918620} of its surroundings with no extra sensing hardware by transforming raw angle delay pairs into explicit spatial coordinates, especially useful for multi-bounce components. 
To this end, we introduce a \textbf{two-segment reflection-point optimization} that estimates the transmitter-to-point and point-to-receiver path lengths separately, handles multi-bounce trajectories, and yields a geometrically consistent scene representation. 
When \ac{CSI} is mined for temporal and spatial channel impulse response details, it then becomes possible to create precise \ac{RF} imaging native to communication links can become possible, paving the way for network-embedded localization, blockage prediction, and environment-aware beam management in $6$G ISAC systems.

\textbf{Channel simulators.}  
Site-specific channel simulators are vital for to evaluate the efficacy of the ability to synthesize a model of the physical 3-D world from standard radio signaling protocols used in networks.
NYURay \cite{9013365,8732419}, a site-specific 3-D ray tracer developed at NYU WIRELESS by two generations of graduate students, is an ideal ray tracing engine for \ac{ISAC} imaging owing to its site-specific ray tracing with the very same angular and temporal parameters a sensing algorithm will use.
\textcolor{black}{ 
In fact, NYURay has been calibrated against real-world $28$, $73$ and $142$ GHz \cite{10721240,10279044,10444718} measurements in complex urban scenes at $6.75$ GHz and $16.95$ GHz \cite{ying2025nyu}. Moreover, NYURay's computations of \ac{ToA}, \ac{ZoA}, \ac{AoA}, \ac{ZoD}, \ac{AoD} and power levels are based on true physics, namely specular reflections, diffuse scattering, penetration loss, and multi-bounce paths \cite{ying2025gc}.} 
Moreover, unlike purely statistical or narrowband solvers, NYURay's hybrid shooting-bouncing rays \cite{10721240} image-based approach handles multiple interactions and large bandwidths simultaneously, yielding per-ray propagation delays, \acp{AoA}/\acp{AoD} and powers across hundreds of MHz or even GHz of bandwidth in a single pass.
NYURay’s deterministic channel model \cite{10579941} replicates exactly the MPCs an ISAC imaging algorithm requires. Consequently, the entire imaging workflow can be prototyped, optimized, and validated in software hence clearing the way for realistic digital-twin simulations.
For large-scale simulation studies requiring system or link-level testing, statistical coverage/capacity analysis, \ac{MIMO} and beamforming evaluation \cite{10901856}, an interesting simulator is NYUSIM \cite{10367974,7996792} which can provide broader statistical insights because of the stochastic channel simulator NYUSIM provides.
NYURay can also be calibrated on recent channel measurements such as $6.75$ GHz \cite{10605910,Ying2025UpperMidBand}.

\textbf{How ray tracing and CSI fit together.}
In \cite{seidel1992tap}, it was first demonstrated that ray-tracing techniques could accurately predict $914$ MHz pathloss in multifloored office environments. Note that the world's first ray tracing dates back to $1991$\cite{schaubach1991propagation}. Indeed, \cite{seidel1992tap} introduced a comprehensive framework for in-building personal communication system design that incorporated deterministic ray interactions, yielding path-specific delay and power predictions that matched measurements to within a few decibels \cite{seidel1994tvt,seidel1993phdthesis}. This body of work established ray tracing as a practical alternative to purely empirical models and directly motivated later advances such as the building-database manipulator \cite{rappaport2005BDM} and reception-surface acceleration techniques \cite{rappaport2004RTsurface}.
Ray tracing and \ac{CSI} operate at complementary layers of the same physical phenomenon. 
A calibrated electromagnetic ray tracer such as NYURay takes an object model, and exact transmitter (TX) and receiver (RX) locations.
Then, ray tracing estimates Maxwell's equations at a finite number of specific locations  by using a large number of emitted radials, called ``rays" \cite{wang2005parametric} from a specified transmitter location within a computerized 3-D environment, and applies basic propagation laws such as reflection, diffraction and scattering \cite{8761205} to each ray. Using the concept of superposition, all rays which arrive at a specified receiver location within the computerized 3-D environment are summed in both amplitude and phase to determine the E-field at the specific location, instead of solving for the full E-field equations.
Ray-tracing was first confirmed to be remarkably accurate for estimating the channel impulse response for wireless networks in \cite{507111,seidel1994tvt,245274}.
The multipath parameters form a “digital twin” of the propagation channel: they describe what the wireless channel must look like before any hardware imperfections or noise. CSI, in contrast, is the measured transfer function obtained in real time by the radios; it embeds the same multipath structure, but convolved with hardware response, clock offsets, and thermal noise. 
We can synthesize CSI snapshots whose per-path geometry is perfectly known, via feeding NYURay's ground-truth rays for channel generation. 
\emph{The synthetic-but-realistic CSI allows us to design and test imaging algorithms, under thousands of TX/RX viewpoints, and when validated, the very same algorithms operate online using live CSI information, turning everyday traffic frames into environmental maps without extra sensing hardware.}


\subsection{Main contributions}
Our work uses \ac{FR3} ray tracing channel measurements provided by NYURay for \ac{ISAC} imaging. In particular, we leverage NYURay's \ac{CSI} to represent multi-bounce scatterers for complex objects.
We have summarized our contributions as follows.
\begin{itemize} \item \textbf{Real data-driven FR3 imaging.} 
\textcolor{black}{We generate NYURay paths for multiple Tx, Rx, and object types at $6.75$ GHz. We feed the NYURay paths information directly into the imaging pipeline, which guarantees \ac{CSI} information that captures genuine material interactions, penetration, diffraction, and diffuse scattering, instead of relying on idealized models.}

\item \textbf{Optimization for equivalent reflection points and Multi-vantage fusion.} We cast the \ac{ISAC} multi-bounce image reconstruction as an optimization problem so as to find the so-called \textit{\textcolor{black}{\acp{ERP}}}, in order to represent objects in the channel of different shapes and sizes,  using transmitter-segment and receiver-segment paths. The segments are given in closed-form.
 In other words, the scattering points are fitted \emph{as if} the reflectors experienced a single-bounce scattering scenario. We jointly recover specular paths and higher-order multipath without the need to assume a single-bounce geometry. Next, in order to build a coherent $3$D reconstruction, we aggregate the \textcolor{black}{\acp{ERP}} from multiple TX-RX vantage pairs. Using a spatial-consistency framework, we fuse the \textcolor{black}{\acp{ERP}} into a single dense point cloud, then apply geometric filtering to reject outliers in order to reliably represent objects, such as trees and vehicles. 
The thresholding is based on how transmitter-segment and receiver-segment deviate from one another. 
Subsequently, we apply an \textcolor{black}{\ac{MVF}} step to provide a complete \ac{RF} image of the object in question.
\end{itemize}

\subsection{Organization \& Notation}

The following paper is organized as follows: 
Section \ref{sec:system-model} presents the equivalent system model adopted in the paper.
Moreover, Section \ref{sec:imaging} formulates an optimization problem tailored to deduce the \textcolor{black}{\acp{ERP}} of the \ac{MPC} components per TX-RX pair. 
Section \ref{sec:simulations} presents the simulation results by showing the resulting images produced by the proposed algorithm.
Section \ref{sec:open-challenges} discusses open challenges regarding \ac{ISAC} imaging.
We conclude the paper in Section \ref{sec:conclusion}.

%
%
%
%
%
%
%
%

\textbf{Notation}: Upper-case and lower-case boldface letters denote matrices and vectors, respectively. $(.)^T$, $(.)^*$ and $(.)^H$ represent the transpose, the conjugate and the transpose-conjugate operators. 
The set of all complex-valued $N \times M$ matrices is $\mathbb{C}^{N \times M}$.  
The Kronecker product is $\otimes$.
\textcolor{black}{The cardinality of set $S$ is denoted as $\left| S \right|$.}
All other notations are defined within the paper.


\textcolor{black}{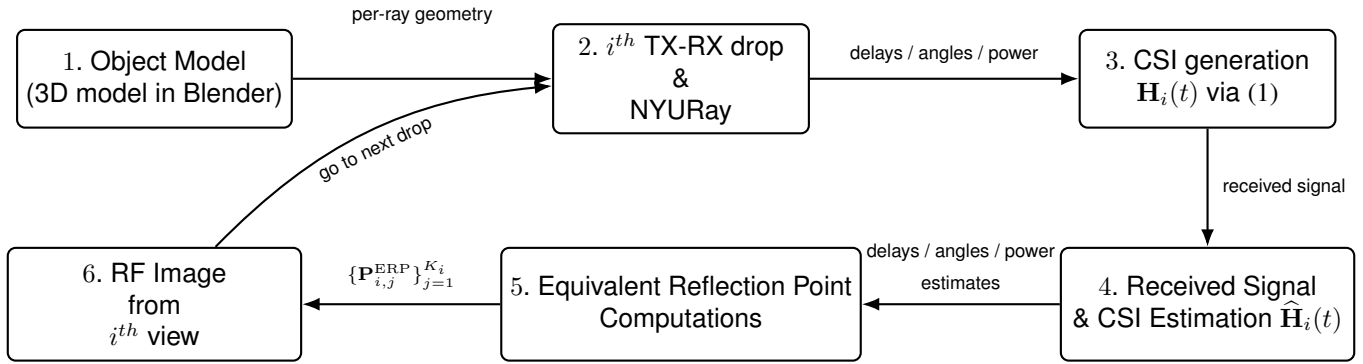
\begin{figure*}[t]
  \centering
  \begin{tikzpicture}[>=Latex, thick, font=\sffamily]

    \node[draw, rounded corners=3pt,
          minimum width=3.4cm, minimum height=1.3cm,
          align=center] (cad) at (0,0)
          {$1$. Object Model\\(3D model in Blender)};

    \node[draw, rounded corners=3pt,
          minimum width=3.4cm, minimum height=1.3cm,
          align=center] (rt) at (7,0)
          {$2$. $i^{th}$ TX-RX drop \\ \& \\ NYURay};

    \node[draw, rounded corners=3pt,
          minimum width=3.4cm, minimum height=1.3cm,
          align=center] (csi) at (14,0)
          {$3$. CSI generation\\$\mathbf{H}_i(t)$ via \eqref{eq:system-model}};


    \node[draw, rounded corners=3pt,
          minimum width=3.9cm, minimum height=1.5cm,
          align=center] (rx)  at (14,-3)
          {$4$. Received Signal\\\& CSI Estimation $\widehat{\mathbf{H}}_i(t)$ };
          
    \node[draw, rounded corners=3pt,
          minimum width=3.4cm, minimum height=1.5cm,
          align=center] (img) at (7,-3)
          {$5$. Equivalent Reflection Point\\ Computations};
          
    \node[draw, rounded corners=3pt,
          minimum width=3.9cm, minimum height=1.5cm,
          align=center] (rx2)  at (0,-3)
          {$6$. RF Image \\ from \\ $i^{th}$ view};

    \draw[->] (cad) -- (rt)
        node[midway, above, yshift=2pt,
          minimum width=3.9cm, minimum height=1.5cm]
        {\scriptsize per-ray geometry};

    \draw[->] (rt) -- (csi)
        node[midway, above, yshift=2pt]
        {\scriptsize delays / angles / power};

    \draw[->] (csi) -- (rx)
        node[midway, right, xshift=2pt]
        {\scriptsize received signal};

\draw[->] (rx) -- (img)
      node[midway, above, yshift=2pt, align=center]
      {\scriptsize delays / angles / power\\ \scriptsize estimates};

    \draw[->] (img) -- (rx2)
        node[midway, above, yshift=2pt]
        {\scriptsize $\lbrace \mathbf{P}^{\mathrm{ERP}}_{i,j} \rbrace_{j=1}^{K_i}$};
        
   \draw[->, bend left=20] (rx2) to
      node[midway, below, sloped]{\scriptsize go to next drop} (rt);
      
  \end{tikzpicture}
  \caption{\textcolor{black}{NYURay-to-CSI imaging pipeline.}}
  \label{fig:rt2csi}
  \vspace{-1.2em}
\end{figure*}}


\section{System Model}
\label{sec:system-model}
\subsection{Mathematical Description}
\label{sec:mathematical-description}
We model the channel at $6.75$ GHz as a superposition of \(K_i\ge1\) path components between the $i^{th}$ TX-RX pair and the object. The \ac{LoS} path, if present, is indexed by \(j=0\), while \(j>0\) denotes the paths that interact with the object in the channel.  
\ac{NLOS} components are generally weaker than the \ac{LOS} path and cannot be assumed to arise from simple \textcolor{black}{single-bounce} reflections, even from objects in the channel. 
\textcolor{black}{
Each path is then characterized by the following six-tuple sensing parameters: 
$\bigl\{\theta_{\mathrm{TX},i,j},\phi_{\mathrm{TX},i,j},
        \theta_{\mathrm{RX},i,j},\phi_{\mathrm{RX},i,j},
        \tau_{i,j},g_{i,j}\bigr\}$, which represent the \ac{AoD}, \ac{ZoD}, \ac{AoA}, \ac{ZoA}, \ac{ToA}, and path gain of the $j^{th}$ path for the $i^{th}$ TX-RX pair, respectively.}
Under the narrowband array assumption, the \ac{MIMO} baseband impulse response is
\textcolor{black}{
\begin{equation}
\label{eq:system-model}
\begin{split}
\mathbf{H}_{i}(t)
&= \sum_{j=0}^{K_i-1}
   g_{i,j}\,
   \mathbf{a}_{\mathrm{RX}}\bigl(\theta_{\mathrm{RX},i,j},\phi_{\mathrm{RX},i,j}\bigr)\,
   \mathbf{a}_{\mathrm{TX}}^T\bigl(\theta_{\mathrm{TX},i,j},\phi_{\mathrm{TX},i,j}\bigr) \\
&\quad \times \delta\bigl(t - \tau_{i,j}\bigr),
\end{split}
\end{equation}
}
\textcolor{black}{In \eqref{eq:system-model}, the transmit and receive array steering vectors are represented as $\mathbf{a}_{\mathrm{TX}}(\theta,\phi)$ and $\mathbf{a}_{\mathrm{RX}}(\theta,\phi)$, respectively.}
\textcolor{black}{As an example, the $3$D steering vector can be written as the Kronecker product of two uniform linear array steering vectors along the \(x\) and \(y\) axes as }\textcolor{black}{
\begin{equation}
\label{eq:cirmodel}
\mathbf{a}_{x}(\theta,\phi)
= \frac{1}{\sqrt{M_x}} \bigl[\ 1\
\ldots\,
e^{-j\frac{2\pi}{\lambda}(M_x-1)d_x\sin\phi\cos\theta}
\bigr]^{T},
\end{equation}
\begin{equation}
\label{eq:cirmodel2}
\mathbf{a}_{y}(\theta,\phi)
= \frac{1}{\sqrt{M_y}}  \bigl[\,1
\ldots\,
e^{-j\frac{2\pi}{\lambda}(M_y-1)d_y\sin\phi\sin\theta}
\bigr]^{T},
\end{equation}
} 
and 
\textcolor{black}{
\begin{equation}
	\mathbf{a}_{\mathrm{RX}/\mathrm{TX}}(\theta, \phi)=\mathbf{a}_y(\theta, \phi) \otimes \mathbf{a}_x(\theta, \phi),
\end{equation}
}
We emphasize the importance of allowing multiple reflections and scattering interactions, which implies that the usual geometric relation between the transmit angles $\bigl(\theta_{\mathrm{TX},i,j},\phi_{\mathrm{TX},i,j}\bigr)$, the receive angles $\bigl(\theta_{\mathrm{RX},i,j},\phi_{\mathrm{RX},i,j}\bigr)$, and the propagation delay $\tau_{i,j}$ no longer holds as it would under a \textcolor{black}{single-bounce} assumption. Therefore, any mismatch among the estimated angles and delays therefore reflects the inherent multi-interaction nature of the propagation environment.
In the rest of the paper, we assume full channel impulse response knowledge of the components of $\mathbf{H}_{i}(t)$, as the channel estimation counterpart is not the main topic of the paper.

\textcolor{black}{Throughout this paper, a \emph{multi-bounce path} denotes any propagation
trajectory in which the electromagnetic wave undergoes two or more
interactions, which may include specular reflection, diffuse scattering, edge diffraction or a mixture thereof, inside the same object before it reaches the receiver.
Explicitly modeling that chain of events would require knowledge of the object's detailed geometry and material parameters, in addition to frequency-dependent electromagnetic properties, in which are unknown a priori in many cases.
Instead, we abstract the entire sequence into a single \textcolor{black}{\ac{ERP}}, which does not attempt to recover the intermediate bounce locations; rather, it captures an equivalent delay and direction change caused by the whole sequence, so that the subsequent imaging step can still rely on a single-bounce geometrical model. In this sense, multi-bounce interactions are \emph{absorbed} into the ERP, eliminating the need to trace or analyze each individual reflection within the object while preserving all information that is observable from the measured data.
}

\subsection{Ray-tracer and CSI}
\label{subsec:raytrace}
In both LTE \cite{TS36211_v1290} and 5G new radio (NR) \cite{TS38211_v1620} the term
CSI refers to the complex MIMO channel-frequency response obtained after correlating the received {CSI-RS} (downlink) or {SRS} (uplink) resource elements with their known pilot sequences $\mathbf{H}[k]$, where each sub-carrier index $k$ contains the channel for every TX-RX antenna pair.
The beamforming header, realized as the swept CSI-RS/SRS resource sets in 5G NR and the BRP-TRN training fields in IEEE 802.11ad/ay, relies on a predefined beam codebook that already delivers a coarse AoD/AoA estimate. Then, to refine the multipath component parameters, $\bigl\{\theta_{\mathrm{TX},i,j},\phi_{\mathrm{TX},i,j},
        \theta_{\mathrm{RX},i,j},\phi_{\mathrm{RX},i,j},
        \tau_{i,j},g_{i,j}\bigr\}$, high-resolution parameter extraction methods can be utilized. 
\textcolor{black}{Indeed, many algorithms exist in the state-of-the-art that observe a received signal (or \ac{CSI}), estimate the propagation channel through the \ac{LMMSE} channel estimator for \ac{OFDM} described in \cite{6814271}}, and produce the sensing parameters $\bigl\{\theta_{\mathrm{TX},i,j},\phi_{\mathrm{TX},i,j},
        \theta_{\mathrm{RX},i,j},\phi_{\mathrm{RX},i,j},
        \tau_{i,j},g_{i,j}\bigr\}$. For example, \cite{7511158} utilizes a matrix pencil approach to estimate the delays and the \acp{AoA}. Deep learning can also be used to estimate the \acp{AoA} and \acp{AoD} \cite{10496165}.

\textcolor{black}{Equations~\eqref{eq:cirmodel} and \eqref{eq:cirmodel2} are} agnostic to where the
per-path parameters come from. 
In this paper, we \emph{generate} the
six-tuple sensing parameters with the NYURay deterministic ray tracer. In essence, 
given a 3D model, NYURay returns every six-tuple multipath component
$\bigl\{\theta_{\mathrm{TX},i,j},\phi_{\mathrm{TX},i,j},
        \theta_{\mathrm{RX},i,j},\phi_{\mathrm{RX},i,j},
        \tau_{i,j},g_{i,j}\bigr\}$ corresponding to the channel between TX and RX in the $i^{th}$ drop, as shown at the output of step 2 in Fig.  \ref{fig:rt2csi}, which are generated based on physics that dictate how each path should interact with the object in question.
\textcolor{black}{Note that the 3D model is rendered on Blender \cite{blender}, which is an open source 3D creation software that enables the modeling, sculpting, and rendering of complex 3D models in one integrated workspace.
On the other hand, a "TX-RX drop" is a terminology indicating a given position of a TX-RX pair and different drops indicate different positions.}        
\textcolor{black}{Following Fig. \ref{fig:rt2csi}, after generating the channel, the received signal can be used to estimate the channel and produce \ac{CSI} information, where the \ac{CSI} can also be used to estimate the parameters $\bigl\{\theta_{\mathrm{TX},i,j},\phi_{\mathrm{TX},i,j},
        \theta_{\mathrm{RX},i,j},\phi_{\mathrm{RX},i,j},
        \tau_{i,j},g_{i,j}\bigr\}$ at RX.}
For each six-tuple path component, an \textcolor{black}{\ac{ERP}} is computed as detailed in Section \ref{sec:imaging}. The process is repeated for each path to form an \ac{RF} image from view $i$ in Step 6 of Fig. \ref{fig:rt2csi}. Once done, the imaging pipeline moves to the next TX-RX drop and repeats the process.
\textcolor{black}{For the proposed imaging technique, we relied on the NYURay engine because of its convenience and calibration capabilities for $6.75$ $\operatorname{GHz}$; however, the proposed technique may be used with other ray tracing tools that provides per-path delays, \ac{ZoD}, \ac{AoD}, \ac{AoA} and \ac{ZoA} information.}\textcolor{black}{It is also worth pointing out that procedures 3 and 4 are shown as conceptual block, and the ground truth paths generated in Procedure 2 are fed directly into Procedure 5, thereby assuming perfect path knowledge.
The actual impact of bandwidth and utilized waveforms are left for future studies as clarified in Section \ref{sec:open-challenges}.}

\textcolor{black}{ In our ray tracer, diffuse scattering is handled by first using material-specific scattering coefficients to decide, at each surface interaction, whether energy is reflected specularly or converted into diffuse rays; when scattering occurs, we generate a set of rays whose directions are sampled uniformly over the hemisphere, assign each a randomized phase to emulate the incoherent nature of diffuse reflections, and scale their amplitudes so that the total scattered energy plus the remaining specular energy equals the incident energy. We further control computational cost by limiting the number of allowed bounces and stochastically pruning very low-power paths. Cross-polarization discrimination is included by splitting scattered energy between orthogonal polarizations according to measured material ratios.
}\textcolor{black}{In this paper, we employ ray tracing simulation to generate synthetic CSI for validating the proposed imaging algorithm capability. Future work will conduct real ISAC experiments with actual ISAC hardware measurements for further validation.}

\section{\textcolor{black}{Imaging via Equivalent Reflection Point Computations}}
\label{sec:imaging}
In our system model, we place the transmitter at $\mathbf{p}_{\mathrm{Tx}_i} \in \mathbb{R}^3$ and the receiver at $\mathbf{p}_{\mathrm{Rx}_i} \in \mathbb{R}^3$. 
The overall path delay of the $(i,j)^{th}$ path delay $\tau_{i,j}$ provides an approximate total path length $L_{i,j} \approx c \tau_{i,j}$.
\textcolor{black}{Using the \ac{ZoD} and \ac{AoD} information, i.e. $\left(\theta_{\mathrm{TX}, i, j}, \phi_{\mathrm{TX}, i, j}\right)$, we denote the unit direction $\hat{\mathbf{d}}_{T_{i,j}}$ as the direction vector emanating from $\mathbf{p}_{\mathrm{Tx}_i}$ towards the object.}
\textcolor{black}{Likewise, we can form the unit direction $\hat{\mathbf{d}}_{R_{i,j}}$ using the \ac{ZoA} and \ac{AoA} information.} 
\textcolor{black}{Following basic geometry}, a scattering point $\mathbf{P}_{i,j}$ along the path must satisfy $\mathbf{P}_{i,j}=\mathbf{p}_{\mathrm{Tx}_i}+\alpha^{(i,j)} \hat{\mathbf{d}}_{T_{i,j}}$ for some $\alpha^{(i,j)}>0$, and $\mathbf{P}_{i,j}=\mathbf{p}_{\mathrm{Rx}_i}-\beta^{(i,j)} \hat{\mathbf{d}}_{R_{i,j}}$ for some $\beta^{(i,j)}>0$. 
\textcolor{black}{The surface on which scattering occurs is denoted by $\mathcal{S} \subset \mathbb{R}^3$, as depicted in Fig. \ref{fig:geom_model}.}
\textcolor{black}{We can think of $\mathbf{a}(\alpha) = \mathbf{p}_{\mathrm{Tx}_i}+\alpha \hat{\mathbf{d}}_{T_{i,j}}$ as the \textit{transmitter-segment} joining the transmitter to the object point scatterer and $\mathbf{b}(\beta) = \mathbf{p}_{\mathrm{Rx}_i}-\beta \hat{\mathbf{d}}_{R_{i,j}}$ can be seen as the \textit{receiver-segment} joining the same object point scatterer to the receiver. The case of $\mathbf{a}(\alpha) = \mathbf{b}(\beta) $ occurs in the case of specular reflection, i.e. when both lines exactly intersect at the scattering point.}

\subsection{\textcolor{black}{Why Directions/Delays are not Geometrically Consistent ?}}
In an idealized single-bounce specular reflection from a perfectly smooth surface, classical geometric optics dictates three strict constraints: 
\textit{(i)} the law of reflection enforces \textcolor{black}{$\angle\left(\mathbf{p}_{\mathrm{Tx}_i} \rightarrow \mathbf{P}_{i,j}\right)=\angle(\mathbf{P}_{i,j} \rightarrow$ $\mathbf{p}_{\mathrm{Rx}_i})$},
\textit{(ii)} the total path length satisfies $\left\|\mathbf{p}_{\mathrm{Tx}_i}-\mathbf{P}_{i,j}\right\|+\left\|\mathbf{P}_{i,j}-\mathbf{p}_{\mathrm{Rx}_i}\right\|=L_{i,j}$, and 
\textit{(iii)} the reflection (or scattering) point $\mathbf{P}_{i,j}$ lies on the surface. However, for diffuse scattering events, where the surface is not perfectly smooth or the scattering occurs over a region rather than a single point, the idealized constraints need not be met simultaneously by any single point $\mathbf{P}_{i,j}$. In practice, measurements (or \textcolor{black}{ray-tracing} outputs) yield \acp{AoD} and \acp{AoA}, along with a \textcolor{black}{propagation delay $\tau_{i,j}$}, \textit{that may define lines in space which do not intersect precisely on the scattering surface.} Consequently, the most accurate approach is to seek a point $\mathbf{P}_{i,j}$ on the surface of the object $\mathcal{S}$ that best approximates the geometric and path-length constraints, rather than enforcing perfect intersection as in the specular reflection model.

\subsection{Formulating an Equivalent Reflection Point Problem}
\textcolor{black}{One method to identifying a scattering point $\mathbf{P}_{i,j} \in \mathcal{S}$ that best fits the geometric and path-length constraints is to formulate the problem in a mathematical optimization framework, which should seek a point $\mathbf{P}_{i,j}$ such that:}
\textit{(i)} the direction from $\mathbf{p}_{\mathrm{Tx}_i}$ to $\mathbf{P}_{i,j}$ \textcolor{black}{approximately} aligns with the estimated \textcolor{black}{\ac{AoD}/\ac{ZoD}}, 
\textit{(ii)} the direction from \textcolor{black}{$\mathbf{P}_{i,j}$} to $\mathbf{p}_{\mathrm{Rx}_i}$ \textcolor{black}{approximately} aligns with the \textcolor{black}{\ac{AoA}/\ac{ZoA}}, 
\textit{(iii)} the sum of the distances \textcolor{black}{$\left\|\mathbf{p}_{\mathrm{Tx}_i}-\mathbf{P}_{i,j}\right\|+\| \mathbf{p}_{\mathrm{Rx}_i}-$ $\mathbf{P}_{i,j} \|$} is \textcolor{black}{approximately corresponds to the propagation delay}, and (iv) $\mathbf{P}_{i,j}$ lies on or very near the scattering surface $\mathcal{S}$.  

\textcolor{black}{\begin{figure}[t]
\centering
\begin{tikzpicture}[>=Latex, thick, font=\small]

\coordinate (Tx)  at (-3.2,  1.0);      
\coordinate (Rx)  at ( 3.2,  1.0);      
\coordinate (A)   at (-0.9,-1.35);      
\coordinate (B)   at ( 0.9,-0.85);      
\coordinate (ERP) at ($ (A)!0.5!(B) $); 

\draw[very thick, smooth, tension=0.8]
      plot coordinates {(-3,-1.3) (-2,-0.9) (-1.2,-1.25)
                        (0,-0.8) (1.2,-0.55) (2,-0.85) (3,-0.6)}
      node[midway, below=5pt, yshift=-20pt,xshift=80pt] {$\mathcal S$};

\draw[fill=black] (Tx) circle(2pt) node[above] {$\mathbf p_{\mathrm{Tx}_i}$};
\draw[fill=black] (Rx) circle(2pt) node[above] {$\mathbf p_{\mathrm{Rx}_i}$};

\draw[fill=black] (A) circle(2pt)
      node[below left=2pt] {$\mathbf{a}\bigl(\alpha^{(i,j)}_{\mathrm{opt}}\bigr)$};

\draw[fill=black] (B) circle(2pt)
      node[below right=2pt] {$\mathbf{b}\bigl(\beta^{(i,j)}_{\mathrm{opt}}\bigr)$};

\draw[fill=white, draw=black, thick] (ERP) circle(2.7pt)
      node[below=1pt] {$\mathbf P^{\mathrm{ERP}}_{i,j}$};

\draw[->] (Tx) -- (A)
      node[midway, above left=1pt, yshift=18pt] {$\alpha^{(i,j)}_{\mathrm{opt}}$};

\draw[->] (B) -- (Rx)
      node[midway, above right=1pt, yshift=12pt] {$\beta^{(i,j)}_{\mathrm{opt}}$};


\end{tikzpicture}
\caption{\textcolor{black}{Geometry of the $(i,j)$-th multipath component. 
The optimization problem in \eqref{eq:original-opt-problem} solves for $\alpha^{(i,j)}_{\mathrm{opt}}$ and $\beta^{(i,j)}_{\mathrm{opt}}$, which aid in generating the \textcolor{black}{\ac{ERP}} $\mathbf{P}^{\mathrm{ERP}}_{i,j}$ following \eqref{eq:ERP-eqn}.
}}
\label{fig:geom_model}
\end{figure}
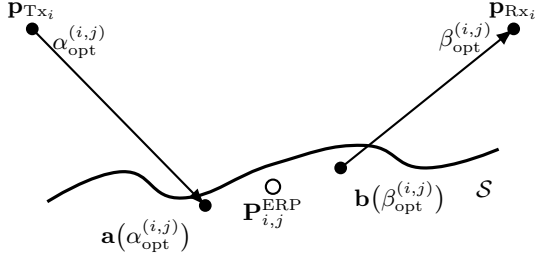

}
Under an ideal single-bounce assumption, the specularly reflected path would satisfy
\begin{equation}
\|\mathbf{p}_{\mathrm{Tx}_i} - \mathbf{P}_{i,j}\| + \|\mathbf{p}_{\mathrm{Rx}_i} - \mathbf{P}_{i,j}\| = L_{i,j},
\end{equation}
which, using the parametric variables \(\alpha^{(i,j)}\) and \(\beta^{(i,j)}\) introduced previously, is equivalent to
\begin{equation}
\alpha^{(i,j)} + \beta^{(i,j)} = L_{i,j}, 
\quad\text{with}\quad \alpha^{(i,j)},\beta^{(i,j)} \ge 0,
\end{equation}
\textcolor{black}{which is imposed to remain consistent with the measured delay \(\tau_{i,j}\).}
Despite the above constraint, there may not exist exact values of \(\alpha^{(i,j)}\) and \(\beta^{(i,j)}\) such that
\begin{equation}
\mathbf{a}(\alpha^{(i,j)}) = \mathbf{b}(\beta^{(i,j)})
\quad \text{and} \quad
\alpha^{(i,j)} + \beta^{(i,j)} = L_{i,j},
\end{equation}
simultaneously. To find the best-fit reflection (or scattering) point, we shall target a least-squares cost to measure the deviation, i.e. 
\begin{equation}
\label{eq:cost-function}
f(\alpha^{(i,j)},\beta^{(i,j)} ) = \bigl\|\mathbf{a}(\alpha^{(i,j)}) - \mathbf{b}(\beta^{(i,j)})\bigr\|^2,
\end{equation}
which represents the squared distance between the two parametric lines.
\textcolor{black}{Next, we state the optimization problem in formal terms as}
\begin{equation}
\label{eq:original-opt-problem}
\left\{
\begin{aligned}
& \min_{\alpha^{(i,j)},\beta^{(i,j)}} 
&& f(\alpha^{(i,j)},\beta^{(i,j)}) = \|\mathbf{a}(\alpha^{(i,j)}) - \mathbf{b}(\beta^{(i,j)})\|^2, \\[6pt]
& \text{subject to}
&& \mathbf{a}(\alpha^{(i,j)}) = \mathbf{p}_{\mathrm{Tx}_i} + \alpha^{(i,j)} \,\hat{\mathbf{d}}_{T_{i,j}}, \\[3pt]
&
&& \mathbf{b}(\beta^{(i,j)}) = \mathbf{p}_{\mathrm{Rx}_i} - \beta^{(i,j)} \,\hat{\mathbf{d}}_{R_{i,j}}, \\[3pt]
&
&& \alpha^{(i,j)} + \beta^{(i,j)} = L_{i,j}, \\[3pt]
&
&& 0 \;\le\; \alpha^{(i,j)}, \beta^{(i,j)} \;\le\; L_{i,j}.
\end{aligned}
\right.	
\end{equation}
\textcolor{black}{The optimization problem in \eqref{eq:original-opt-problem} treats the $j^{th}$ path for the $i^{th}$ TX-RX pair, separately.}
\textcolor{black}{An alternative is to set up a single optimization problem that spans all propagation paths and every TX-RX pair, expressed through the aggregate cost $\sum_i \sum_j f\left(\alpha^{(i, j)}, \beta^{(i, j)}\right)$, or even more a weighted aggregate cost $\sum_i \sum_j\left|g_{i, j}\right|^2 f\left(\alpha^{(i, j)}, \beta^{(i, j)}\right)$ which can be introduced by weighting each term by a reliability according to the path power $\left|g_{i, j}\right|^2$. The aggregate cost problem may be left for future work.}
\textcolor{black}{To construct the Lagrangian for problem (\ref{eq:original-opt-problem}), we associate the Lagrange multiplier $\lambda$ to the equality constraint 
$\alpha^{(i,j)} + \beta^{(i,j)} - L_{i,j} = 0$ 
and assign the multipliers $\mu_1, \mu_2, \mu_3,$ and $\mu_4$ to the inequality constraints 
$\alpha^{(i,j)} \ge 0$, 
$\alpha^{(i,j)} \le L_{i,j}$, 
$\beta^{(i,j)} \ge 0$, 
and 
$\beta^{(i,j)} \le L_{i,j}$, respectively.  
Hence, the constraints can be written as}
\begin{equation}
\label{eq:consitraints-rewritten}
	\begin{aligned} & g_1(\alpha^{(i,j)}, \beta^{(i,j)})=\alpha^{(i,j)}+\beta^{(i,j)}-L_{i,j}=0, \\ & h_1(\alpha^{(i,j)}, \beta^{(i,j)})=-\alpha^{(i,j)} \leq 0,  \\ & h_2(\alpha^{(i,j)}, \beta^{(i,j)})=\alpha^{(i,j)}-L_{i,j} \leq 0, \\ & h_3(\alpha^{(i,j)}, \beta^{(i,j)})=-\beta^{(i,j)} \leq 0, \\ &  h_4(\alpha^{(i,j)}, \beta^{(i,j)})=\beta^{(i,j)}-L_{i,j} \leq 0 .\end{aligned}
\end{equation}
\textcolor{black}{We can now write the corresponding Lagrangian as}
\begin{equation}
	\begin{split}
	&\mathcal{L}\left(\alpha^{(i,j)}, \beta^{(i,j)}, \lambda, \mu_1, \mu_2, \mu_3, \mu_4\right) \\ &=\|\mathbf{a}(\alpha^{(i,j)})-\mathbf{b}(\beta^{(i,j)})\|^2+\lambda(\alpha^{(i,j)}+\beta^{(i,j)}-L_{i,j})\\ &+\mu_1(-\alpha^{(i,j)})+\mu_2(\alpha^{(i,j)}-L_{i,j}) \\ &+\mu_3(-\beta^{(i,j)})+\mu_4(\beta^{(i,j)}-L_{i,j}).
	\end{split}
\end{equation}
\textcolor{black}{To establish the stationarity conditions, we first differentiate $\mathcal{L}$ with respect to $\alpha^{(i,j)}$ and $\beta^{(i,j)}$. Observe that the cost term in \eqref{eq:cost-function} can be rewritten in quadratic form as $f\left(\alpha^{(i, j)}, \beta^{(i, j)}\right)=\mathbf{v}^T\left(\alpha^{(i, j)}, \beta^{(i, j)}\right) \mathbf{v}\left(\alpha^{(i, j)}, \beta^{(i, j)}\right)$ where the vector $\mathbf{v}(\alpha^{(i,j)}, \beta^{(i,j)})$ is given by}
\begin{equation}
\label{eq:v_equation}
	\mathbf{v}(\alpha^{(i,j)}, \beta^{(i,j)})=\mathbf{p}_{\mathrm{Tx}_i}- \mathbf{p}_{\mathrm{Rx}_i} +\alpha^{(i,j)} \hat{\mathbf{d}}_{T_{i,j}}+ \beta^{(i,j)} \hat{\mathbf{d}}_{R_{i,j}}.
\end{equation}
The gradients of $f$ with respect to $\alpha^{(i,j)}$ and $\beta^{(i,j)}$ are then
\begin{equation}
\label{eq:nabla_a}
	\nabla_{\alpha^{(i,j)}} f(\alpha^{(i,j)}, \beta^{(i,j)})=2 \mathbf{v}(\alpha^{(i,j)}, \beta^{(i,j)})\hat{\mathbf{d}}_{T_{i,j}},
\end{equation}
\begin{equation}
\label{eq:nabla_b}
	\nabla_{\beta^{(i,j)}} f(\alpha^{(i,j)}, \beta^{(i,j)})=2 \mathbf{v}(\alpha^{(i,j)}, \beta^{(i,j)}) \hat{\mathbf{d}}_{R_{i,j}}.
\end{equation}
Inserting \eqref{eq:nabla_a} and \eqref{eq:nabla_b} into the Lagrangian $\mathcal{L}\left(\alpha^{(i,j)}, \beta^{(i,j)},\textcolor{black}{ \lambda,\left\{\mu_k\right\}}\right)$ yields the stationarity conditions
\begin{equation}
\begin{aligned} & \frac{\partial \mathcal{L}}{\partial \alpha^{(i,j)}}=2 \mathbf{v}(\alpha^{(i,j)}, \beta^{(i,j)})  \hat{\mathbf{d}}_{T_{i,j}}+\lambda-\mu_1+\mu_2=0, \\ & \frac{\partial \mathcal{L}}{\partial \beta^{(i,j)}}=2 \mathbf{v}(\alpha^{(i,j)}, \beta^{(i,j)})  \hat{\mathbf{d}}_{R_{i,j}}+\lambda-\mu_3+\mu_4=0.
\end{aligned}	
\end{equation}
\textcolor{black}{Under the \ac{KKT} complementary slackness condition, every inequality constraint \textcolor{black}{$h_k\left(\alpha^{(i, j)}, \beta^{(i, j)}\right) \leq 0$ }defined in \eqref{eq:consitraints-rewritten} must be either inactive, with its multiplier equal to zero, or active, in which case the associated multiplier adjusts to satisfy the condition.}
Formally, for \textcolor{black}{$\mu_k \geq 0$} and \textcolor{black}{$h_k(\alpha^{(i,j)}, \beta^{(i,j)}) \leq 0$}, the condition \textcolor{black}{$\mu_k h_k(\alpha^{(i,j)}, \beta^{(i,j)})=0$} must hold. In the optimization problem defined in equation (\ref{eq:original-opt-problem}), the \ac{KKT} conditions translate to
\begin{subequations}
	\begin{equation}
		\mu_1 \geq 0, \quad \mu_1 (-\alpha^{(i,j)})=0,
	\end{equation}
	\begin{equation}
		\mu_2 \geq 0, \quad \mu_2 (\alpha^{(i,j)}-L_{i,j})=0,
	\end{equation}
	\begin{equation}
		\mu_3 \geq 0, \quad \mu_3 (-\beta^{(i,j)})=0,
	\end{equation}
	\begin{equation}
		\mu_4 \geq 0, \quad \mu_4 (\beta^{(i,j)}-L_{i,j})=0.
	\end{equation}
\end{subequations}
Since $0<\alpha^{(i,j)}<L_{i,j}$ and $0<\beta^{(i,j)}<L_{i,j}$, then all inequality constraints are inactive, so
\begin{equation}
	\mu_1=\mu_2=\mu_3=\mu_4=0.
\end{equation}
\textcolor{black}{Thus, the only active constraint is $\alpha^{(i, j)}+\beta^{(i, j)}=L_{i, j}$; therefore, $\lambda \neq 0$ is obtained from the stationarity conditions, and the problem simplifies to}
\begin{subequations}
	\begin{equation}
		\alpha^{(i,j)}+\beta^{(i,j)}=L_{i,j},
	\end{equation}
	\begin{equation}
	\label{eq:nabla_a_plus_lamba}
		\nabla_{\alpha^{(i,j)}} f(\alpha^{(i,j)}, \beta^{(i,j)})+\lambda=0,
	\end{equation}
	\begin{equation}
	\label{eq:nabla_b_plus_lamba}
		\nabla_{\beta^{(i,j)}} f(\alpha^{(i,j)}, \beta^{(i,j)})+\lambda=0.
	\end{equation}
\end{subequations}
\textcolor{black}{After substituting \eqref{eq:nabla_a} and \eqref{eq:nabla_b} into \eqref{eq:nabla_a_plus_lamba} and \eqref{eq:nabla_b_plus_lamba}, respectively, and subtracting the two resulting expressions, we obtain}
\begin{equation}
\label{eq:after_subt}
	2\left[\mathbf{v}(\alpha^{(i,j)}, \beta^{(i,j)})  \hat{\mathbf{d}}_{T_{i,j}}-\mathbf{v}(\alpha^{(i,j)}, \beta^{(i,j)})  \hat{\mathbf{d}}_{R_{i,j}}\right]=0
\end{equation}
\textcolor{black}{Invoking \eqref{eq:v_equation} and rewriting $\mathbf{v}\left(\alpha^{(i, j)}, \beta^{(i, j)}\right)$ in terms of $\alpha^{(i, j)}$ by setting $\beta^{(i, j)}=L-\alpha^{(i, j)}$, hence $\mathbf{v}\left(\alpha^{(i, j)}, L-\alpha^{(i, j)}\right)=\left(\mathbf{p}_{\mathrm{Tx}_i}-\mathbf{p}_{\mathrm{Rx}_i}\right)+\alpha^{(i, j)} \hat{\mathbf{d}}_{T_{i, j}}+\left(L-\alpha^{(i, j)}\right) \hat{\mathbf{d}}_{R_{i, j}}$, equation \eqref{eq:after_subt} becomes}
\textcolor{black}{
\begin{equation}
\label{eq:solve-for-alpha}
\begin{split}
\bigl[(\mathbf{p}_{\mathrm{Tx}_i}-\mathbf{p}_{\mathrm{Rx}_i})
      +L\hat{\mathbf{d}}_{R_{i,j}}
      +\alpha^{(i,j)}\!\bigl(\hat{\mathbf{d}}_{T_{i,j}}-\hat{\mathbf{d}}_{R_{i,j}}\bigr)\bigr]^{\!T}\\
\cdot\bigl(\hat{\mathbf{d}}_{T_{i,j}}-\hat{\mathbf{d}}_{R_{i,j}}\bigr)=0.
\end{split}
\end{equation}
}
Hence the optimal $\alpha^{(i,j)}$ which solves \eqref{eq:solve-for-alpha} is 
\begin{equation}
\label{eq:one-ERP-computation}
	\alpha^{(i,j)}_{\mathrm{opt}}=-\frac{[\mathbf{p}_{\mathrm{Tx}_i}-\mathbf{p}_{\mathrm{Rx}_i}+L \hat{\mathbf{d}}_{R_{i,j}}]^T \left[\hat{\mathbf{d}}_{T_{i,j}}-\hat{\mathbf{d}}_{R_{i,j}}\right]}{\left(\hat{\mathbf{d}}_{T_{i,j}}-\hat{\mathbf{d}}_{R_{i,j}}\right) ^T\left(\hat{\mathbf{d}}_{T_{i,j}}-\hat{\mathbf{d}}_{R_{i,j}}\right)},
\end{equation}
and therefore $\beta^{(i,j)}_{\mathrm{opt}}=L_{i,j}-\alpha^{(i,j)}_{\mathrm{opt}}$.
The solution is for \ac{NLOS} components, where $j > 0$, as per the definition.
Finally, if $0 \leq \alpha^{(i,j)}_{\mathrm{opt}} \leq L_{i,j} $, then select $\alpha^{(i,j)}_{\mathrm{opt}}$ as the minimzer (in addition to $\beta^{(i,j)}_{\mathrm{opt}}=L_{i,j}-\alpha^{(i,j)}_{\mathrm{opt}}$) because both satisfy the box-constraint in \eqref{eq:original-opt-problem}. 
Otherwise, if $\alpha^{(i,j)}_{\mathrm{opt}}$ is not within $[0,L_{i,j}]$, then pick $(\alpha^{(i,j)}_{\mathrm{opt}},\beta^{(i,j)}_{\mathrm{opt}})$ in a way to minimize $f(\alpha^{(i,j)},\beta^{(i,j)})$ at the boundary, namely either $(0,L_{i,j})$ or $(L_{i,j},0)$.

The \textcolor{black}{\ac{ERP}} is then given by
\begin{equation}
\label{eq:ERP-eqn}
\mathbf{P}^{\mathrm{ERP}}_{i,j} = \frac{\mathbf{a}(\alpha^{(i,j)}_{\mathrm{opt}}) + \mathbf{b}(\beta^{(i,j)}_{\mathrm{opt}})}{2}.
\end{equation}
\textcolor{black}{For a pure single-bounce path, the cost function vanishes, i.e. $f\left(\alpha_{\mathrm{opt}}^{(i, j)}, \beta_{\mathrm{opt}}^{(i, j)}\right)=0$, so the transmitter-segment and receiver-segment intersect at a unique point.}
However, in the general case, where the path can encounter multiple sub-bounces or a distributed scattering surface, we have that $f(\alpha^{(i,j)}) > 0$. 
\textcolor{black}{In the general case, $\mathbf{P}_{i, j}^{\mathrm{ERP}}$ should be viewed as an effective reflection point, which defines the location that best fits all parameters $\bigl\{\theta_{\mathrm{TX},i,j},\phi_{\mathrm{TX},i,j},
        \theta_{\mathrm{RX},i,j},\phi_{\mathrm{RX},i,j},
        \tau_{i,j},g_{i,j}\bigr\}$. Although the \textcolor{black}{\ac{ERP}} is only a geometric best-fit to a potentially complex physical trajectory, it provides the best-fit location for the dominant interaction associated with each path component, in the least-squares sense.}

\subsection{\textcolor{black}{Multi-vantage fusion}}
\label{sec:MF-Fusion}
\textcolor{black}{After computing all the \textcolor{black}{\acp{ERP}} covering all Tx-Rx pair $\mathbf{P}_{i, j}^{\mathrm{ERP}}$ following equation \eqref{eq:ERP-eqn}, we apply a geometric filtering step and reject all \textcolor{black}{\acp{ERP}}  when $\Vert \mathbf{a}\left(\alpha_{\mathrm{opt}}^{(i, j)}\right)-\mathbf{b}\left(\beta_{\mathrm{opt}}^{(i, j)}\right) \Vert \geq \gamma$. 
The parameter $\gamma$ measures the straight line chord length that links the entry and exit points of that multi-bounce path inside the object.
The role of $\gamma$ is to reject candidate \textcolor{black}{\acp{ERP}} whose incoming and outgoing rays, captured by $\mathbf{a}(\alpha^{(i,j)}_{\mathrm{opt}})$ and $ \mathbf{b}(\beta^{(i,j)}_{\mathrm{opt}})$ would have to diverge by an unreasonable large internal path, typically a sign of poor optimization convergence.
   
To this end, the \textcolor{black}{\ac{MVF}} step aggregates the \textcolor{black}{\acp{ERP}} as follows
\begin{equation}
\label{eq:Sgamma}
\mathcal{S}_\gamma
=
\bigcup_{i=1}^{N_p}
\Bigl\{
  \mathbf{P}^{\mathrm{ERP}}_{i,j}
  :
  \lVert
    \mathbf{a}(\alpha^{(i,j)}_{\mathrm{opt}})
    -
    \mathbf{b}(\beta^{(i,j)}_{\mathrm{opt}})
  \rVert
  <
  \gamma
\Bigr\}.
\end{equation}
Equation~\eqref{eq:Sgamma} employs a fixed Euclidean bound, but the same framework readily
admits more sophisticated criteria.  For instance, one may replace the
hard threshold~$\gamma$ by an \emph{adaptive} radius
$\gamma_{i,j}=c\,\sigma_{i,j}$ proportional to an uncertainty estimate
$\sigma_{i,j}$ obtained from the optimisation residual or the Cram\'er-Rao
bound. One can also consider the Mahalanobis  distance instead 
$\bigl\lVert\mathbf{C}_{i,j}^{-1/2}
  \bigl(\mathbf{a}(\alpha^{(i,j)}_{\mathrm{opt}})
  -\mathbf{b}(\beta^{(i,j)}_{\mathrm{opt}})\bigr)\bigr\rVert$
to account for anisotropic error ellipsoids given prior information of the object to be imaged.}

\subsection{\textcolor{black}{Computational Complexity}}
\label{sec:complexity}
\textcolor{black}{The computational complexity for one \textcolor{black}{\ac{ERP}} computation following equation \eqref{eq:one-ERP-computation} in the updated manuscript is only $\mathcal{O}(1)$ due to the abstraction of intermediate bounces as an approximate \textcolor{black}{\ac{ERP}}. 
Assuming $N$ TX-RX pairs and bounding by the worst-case number of path components as $\max_i K_i$, the overall computational complexity to compute all \textcolor{black}{\acp{ERP}} over all the TX-RX pairs to form a dense three dimensional point cloud is $\mathcal{O}(N \max_i K_i)$, which is linear in $N$.}

\section{Simulation Results}
\label{sec:simulations}
\begin{table}[ht]
\centering
\caption{Key Simulation Parameters}
\begin{tabular}{@{}ll@{}}
\toprule
\textbf{Parameter}                    & \textbf{Value / Assumption}                          \\ \midrule
TX/RX Antenna Pattern                & Isotropic, 0 dBi gain in all directions              \\
Transmit Power                       & 0 dBm                                                \\
Carrier Frequency                    & 6.75 GHz                                             \\
Minimum Received Power    & -160 dBm (CIR cutoff threshold)                      \\
Ray Tracing Mechanisms Enabled       & Reflection, Scattering, Diffraction                  \\
Noise and Bandwidth Constraints      & Not included (idealized CIR used)                   \\ \bottomrule
\end{tabular}
\label{tab:sim_params}
\end{table}

\begin{figure*}[!t]
  \centering
  \subfloat[Standard tree model]{%
    \includegraphics[width=0.49\textwidth]{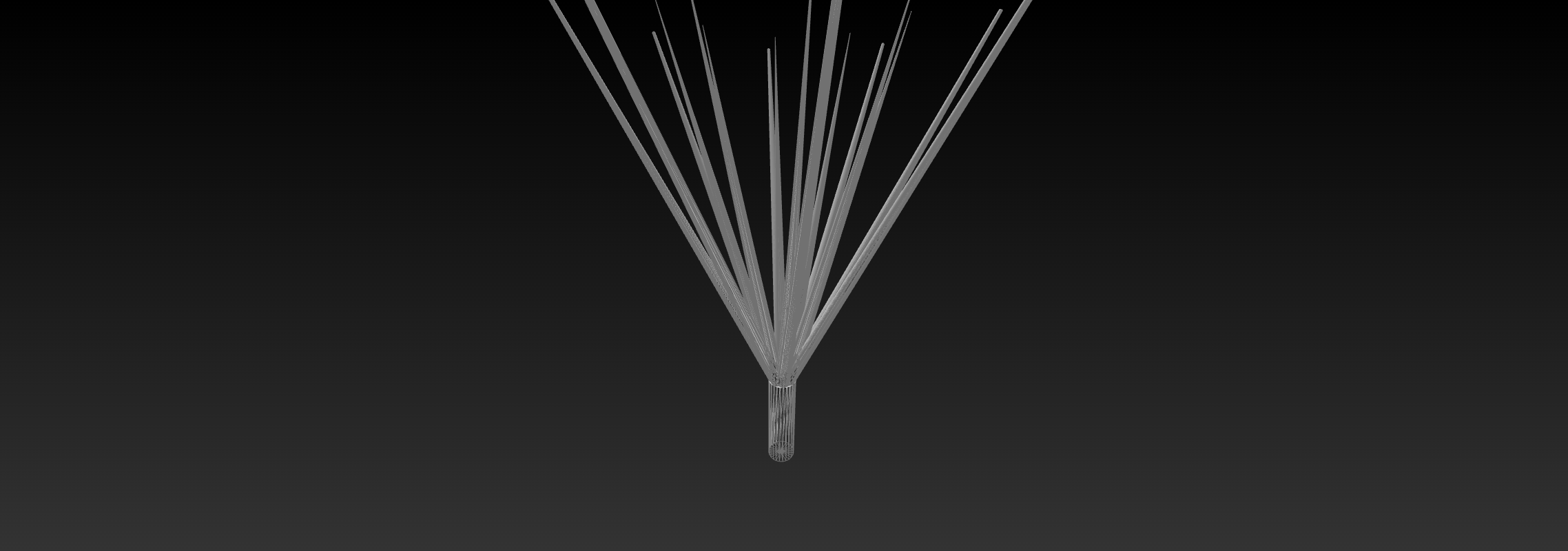}%
    \label{fig:blend1}%
  }\hfill
  \subfloat[Metal Cube (\textcolor{black}{1 m}) model]{%
    \includegraphics[width=0.49\textwidth]{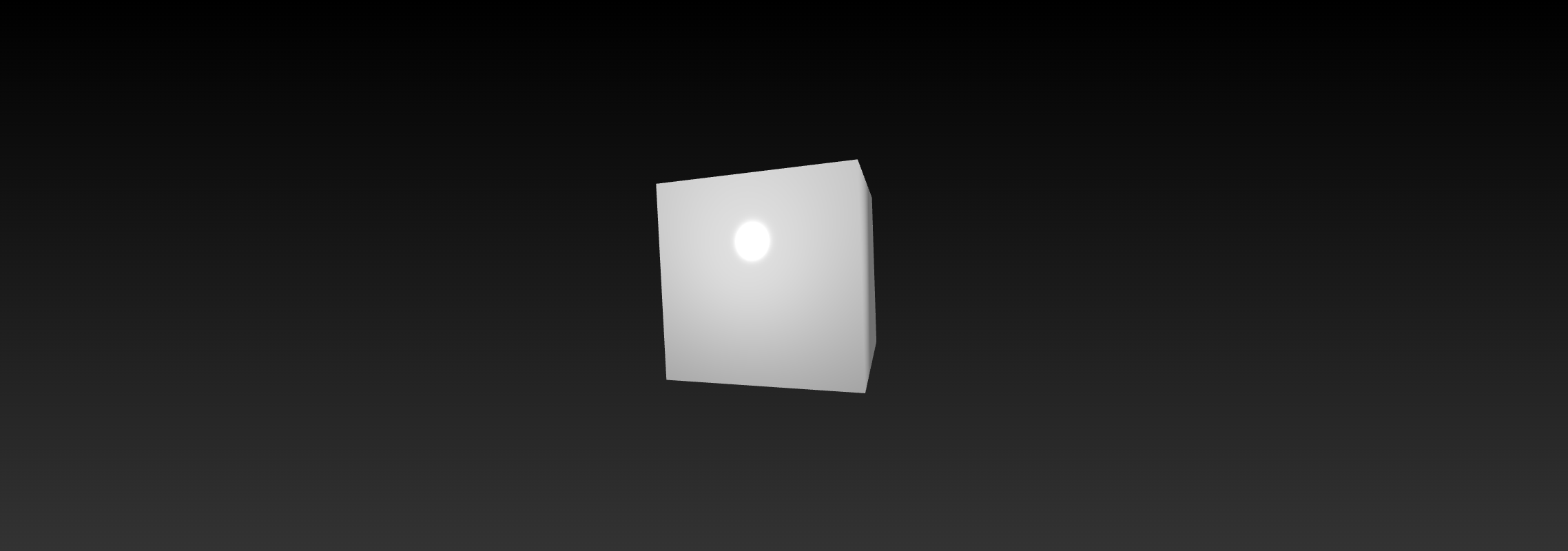}%
    \label{fig:blend2}%
  }\hfill
  \subfloat[Metal Cube (\textcolor{black}{4 m}) model]{%
    \includegraphics[width=0.49\textwidth]{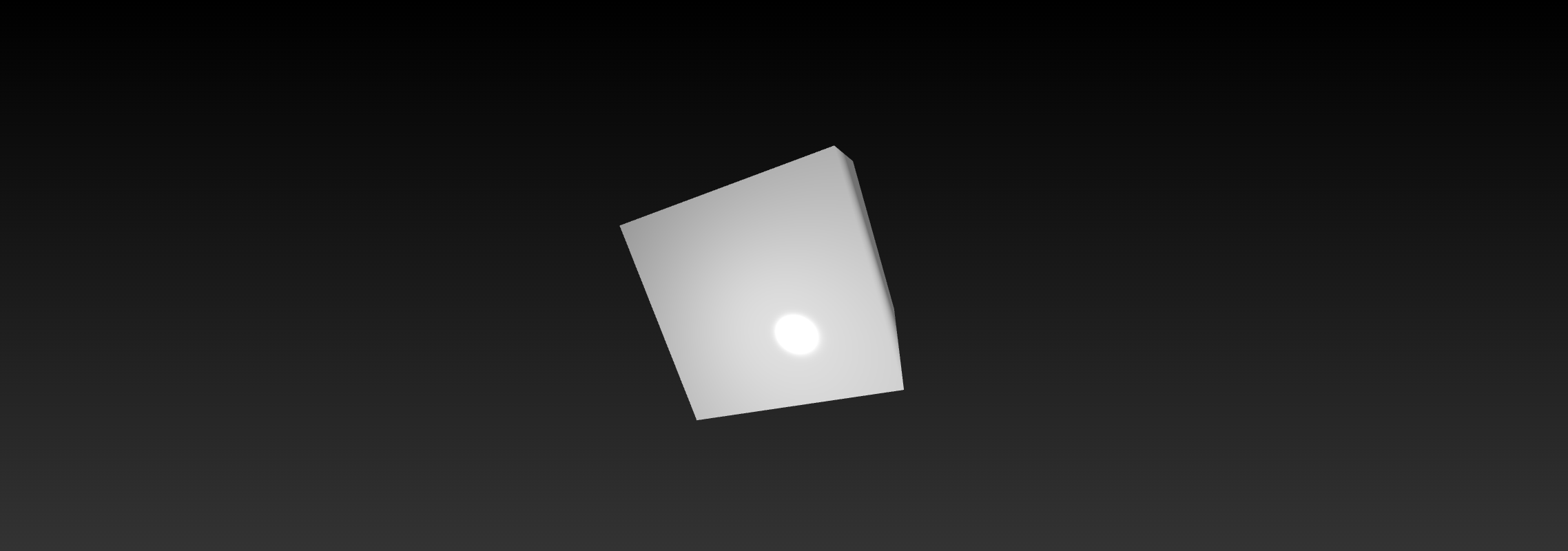}%
    \label{fig:blend3}%
  }\hfill
  \subfloat[Metal Triangle model]{%
    \includegraphics[width=0.49\textwidth]{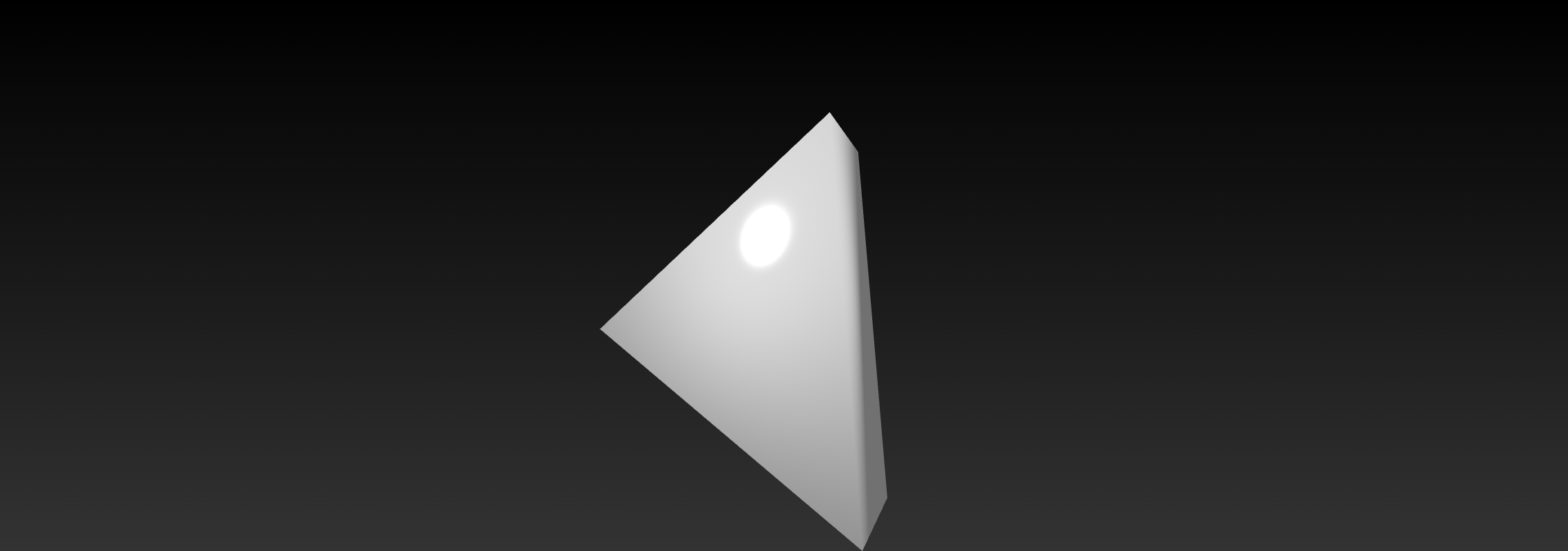}%
    \label{fig:blend4}%
  }\hfill
  \subfloat[Metal Circle model]{%
    \includegraphics[width=0.49\textwidth]{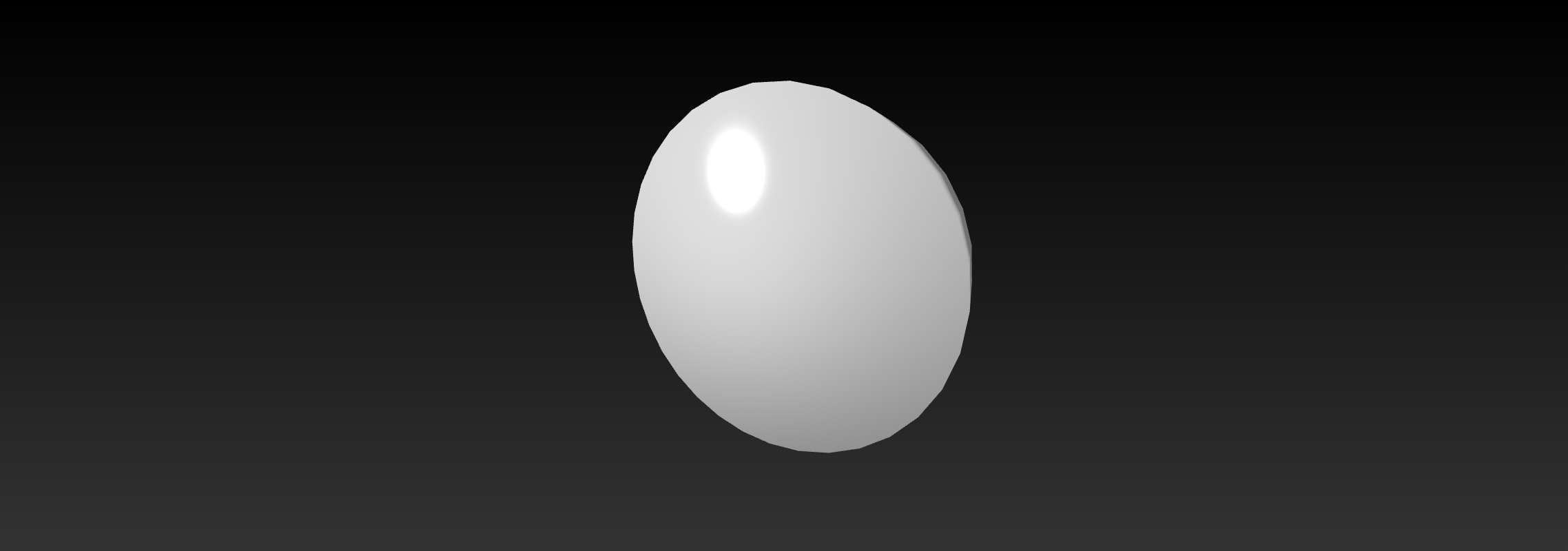}%
    \label{fig:blend5}%
  }\hfill
  \subfloat[Tesla model]{%
    \includegraphics[width=0.49\textwidth]{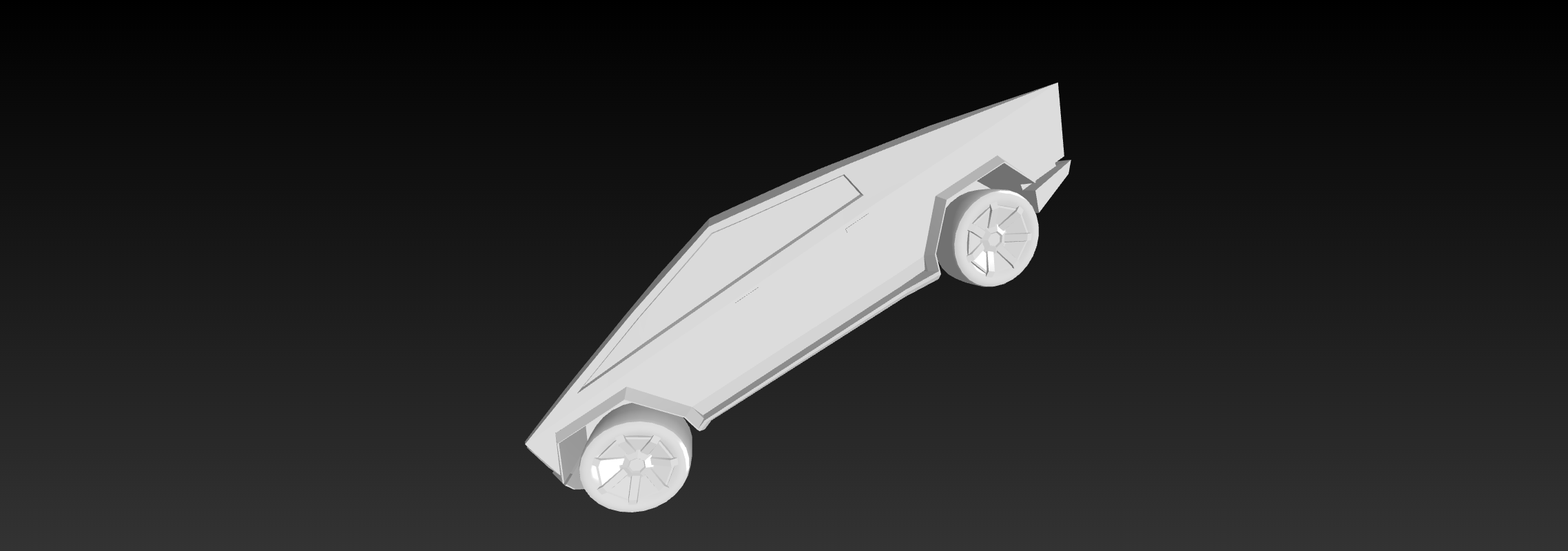}%
    \label{fig:blend6}%
  }\hfill
  \caption{Six 3D models rendered in Blender and used as test objects in our imaging experiments.}
  \label{fig:blend}
\end{figure*}
Throughout our simulations, we evaluated the proposed \ac{ISAC} imaging algorithm on six distinct $3$D test objects rendered in Blender which are given in Fig.~\ref{fig:blend}.
In particular, we selected a standard tree model to assess performance on complex, branched geometries as depicted in Fig.~\ref{fig:blend1}.
\textcolor{black}{The standard tree model used in our simulations represents an averaged structure derived from measurements of multiple real trees, with a trunk diameter of $0.3$ meters, trunk height of $2.1$ meters, and total height of $8$ meters.
The model includes $16$ branches with conical geometry approximations to maintain essential scattering characteristics while reducing computational complexity.
The electromagnetic properties for wood are based on ITU-R P.$2040$-$3$ \cite{ITU-R_P2040-3_2023}, with a constant relative permittivity of $1.99$ and frequency-dependent conductivity $\sigma= 0.0047 \times f^{1.0718}$ S/m (where $f$ is frequency in GHz), valid over the frequency range of $0.001$-$100$ $\operatorname{GHz}$, where the conductivity is taken from ITU-R P.$2040$-$3$ \cite{ITU-R_P2040-3_2023}.
This standardized model enables realistic simulation of dielectric scattering in typical outdoor environments around NYU Brooklyn campus while maintaining computational efficiency for the ISAC framework validation.} In Fig.~\ref{fig:blend2} and Fig.~\ref{fig:blend3}, we have two metal cubes of differing edge lengths, 1 m and 4 m, respectively, in order to characterize specular reflections and scale effects. Moreover, in Fig.~\ref{fig:blend4}, a metal triangular is utilized to probe non‑orthogonal planar facets.
In addition, Fig.~\ref{fig:blend5} represents a smooth metal circle to examine  scattering from curved bodies; and Fig.~\ref{fig:blend6} represents a Tesla body to demonstrate efficacy on a realistic vehicular silhouette. Collectively, the six models span a wide variety of shapes, surface normals and material symmetries, allowing us to comprehensively test the accuracy and robustness of our \ac{ISAC} imaging algorithm under diverse scattering conditions.
\textcolor{black}{The transmit and receive number of antennas are set to $M_x=M_y=1$ isotropic antennas with $0$ dBi gain in all directions. The transmit power is set to $0$ dBm and the carrier frequency is $6.75$ GHz. 
In addition, the minimum receive power specified by the \ac{CIR} cutoff threshold is $-160$ dBm.}
For geometric filtering, we have set $\gamma = 10$. 
\textcolor{black}{We have pick $\gamma = 10$ because the longest straight-line, edge-to-edge dimension of any target in our data set, which exceeds that maximum internal chord.}
Our key simulation parameters are summarized in Table \ref{tab:sim_params}. \textcolor{black}{It is important to recognize that the imaging algorithm is frequency-agnostic, and we have chosen to utilize the calibrated 6.75 GHz NYURay ray tracer based on the vast data set of real-world channel measurements as reported in \cite{Rappaport2025InH,10978475,10605910,10901735,shakya2025urban,Ying2025UpperMidBand,shakya2025upper,bazzi2025upper}. This approach ensures accurate modeling, but in no way limits the generality of the methods described herein.}

\begin{figure*}[!t]
  \centering
  \subfloat[View 1]{%
    \includegraphics[width=0.24\textwidth]{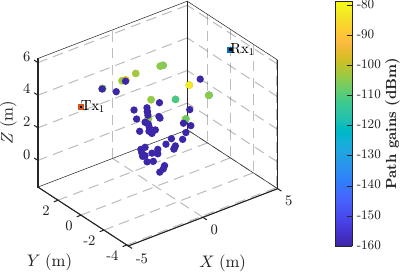}%
    \label{fig:st1}%
  }\hfill
  \subfloat[View 2]{%
    \includegraphics[width=0.24\textwidth]{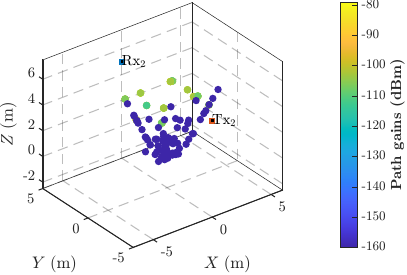}%
    \label{fig:st2}%
  }\hfill
  \subfloat[View 3]{%
    \includegraphics[width=0.24\textwidth]{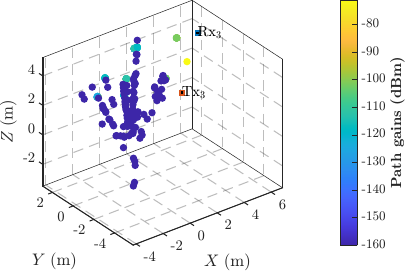}%
    \label{fig:st3}%
  }\hfill
  \subfloat[View 4]{%
    \includegraphics[width=0.24\textwidth]{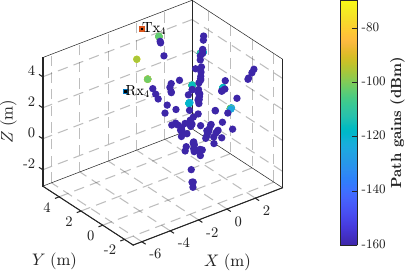}%
    \label{fig:st4}%
  }

  \vspace{0.5em}

  \subfloat[View 5]{%
    \includegraphics[width=0.24\textwidth]{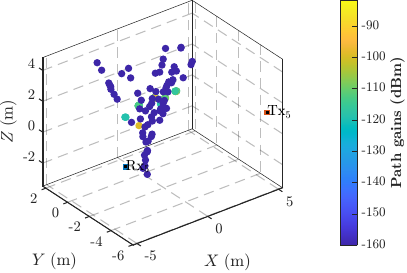}%
    \label{fig:st5}%
  }\hfill
  \subfloat[View 6]{%
    \includegraphics[width=0.24\textwidth]{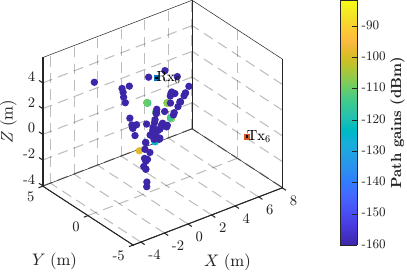}%
    \label{fig:st6}%
  }\hfill
  \subfloat[View 7]{%
    \includegraphics[width=0.24\textwidth]{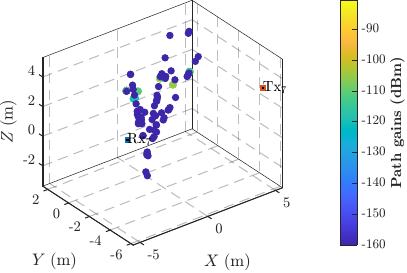}%
    \label{fig:st7}%
  }\hfill
  \subfloat[\textcolor{black}{\ac{MVF}}]{%
    \includegraphics[width=0.24\textwidth]{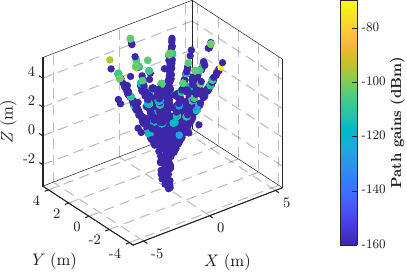}%
    \label{fig:stf}%
  }

  \caption{Eight representative views of the standard tree scenario.}
  \label{fig:standard-tree-collection}
\end{figure*}

Figure~\ref{fig:standard-tree-collection} presents representative 3D reconstructions produced by the proposed imaging algorithm at $6.75$ GHz for the standard tree scenario, whose Blender model is shown in Fig.~\ref{fig:blend1}. 
Each point on the figure represent the \textit{\textcolor{black}{\ac{ERP}}} computed by the proposed framework. 
Each point is colored by its path gain.  
Fig.~\ref{fig:st1} to Fig.~\ref{fig:st7} correspond to seven distinct TX-RX placements, while Fig.~\ref{fig:stf} fuses all seven outputs into a single composite \ac{RF} image.  
\textcolor{black}{The seven distinct placements are given by the rows of the following matrices, where the $i^{th}$ row gives the coordinates of the TX and RX, respectively:
\begin{equation}
\operatorname{Tx}_{\operatorname{pos}}
=
\begin{bmatrix}
-5 &  0 & 5 \\
 0 & -5 & 5 \\
 0 & -5 & 5 \\
 0 &  5 & 5 \\
 5 & -5 & 1 \\
 5 & -5 & 1 \\
 5 & -5 & 3
\end{bmatrix},
\quad
\operatorname{Rx}_{\operatorname{pos}}
=
\begin{bmatrix}
 5 & 0 & 5 \\
 0 & 5 & 5 \\
 5 & 0 & 5 \\
-5 & 0 & 5 \\
-5 & -5 & 1 \\
 5 & 5 & 1 \\
-5 & -5 & 3
\end{bmatrix}.
\end{equation}}
The reconstructions reveal a clear vertical stratification of returns: the strongest reflections concentrate in the mid-canopy band (\textcolor{black}{$Z\approx 1 \ldots 4$ m}) with a marked drop in both point density and recovered gain below \textcolor{black}{$Z=1$ m}, indicating that \textcolor{black}{ground-bounce} and low-branch paths play only a minor role.
Furthermore, each individual view exhibits “shadow zones” behind dense foliage, yet the gaps are filled by complementary perspectives. 
For example, the views in Fig.~\ref{fig:st1} and Fig.~\ref{fig:st2} can recover the edges of tree branches and foliage. On the other hand, the views in Fig.~\ref{fig:st5} and Fig.~\ref{fig:st6} can identify the trunk of the tree.   
Finally, the fused \ac{RF} image in Fig.~\ref{fig:stf} produces an almost continuous shell of scatterers around the mid-canopy, dramatically improving spatial and angular coverage compared to any single view. 
Together, different views validate that the proposed imaging algorithm not only localizes foliage and trunk returns with high fidelity but also when fused across multiple TX-RX configurations, the method delivers a comprehensive 3D map of the standard tree object at $6.75$ GHz.

\begin{figure*}[!t]
  \centering
  \subfloat[View 1]{%
    \includegraphics[width=0.24\textwidth]{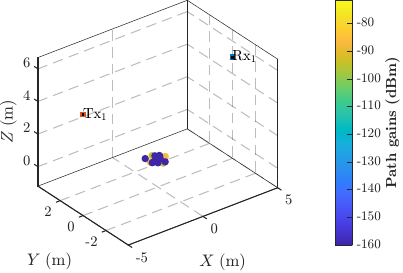}%
    \label{fig:mc1m1}%
  }\hfill
  \subfloat[View 2]{%
    \includegraphics[width=0.24\textwidth]{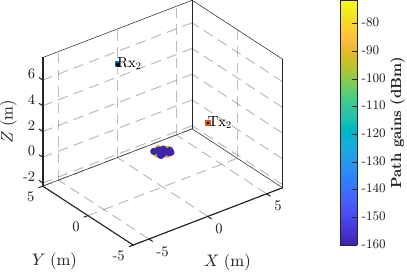}%
    \label{fig:mc1m2}%
  }\hfill
  \subfloat[View 3]{%
    \includegraphics[width=0.24\textwidth]{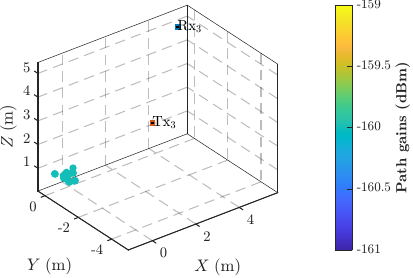}%
    \label{fig:mc1m3}%
  }\hfill
  \subfloat[View 4]{%
    \includegraphics[width=0.24\textwidth]{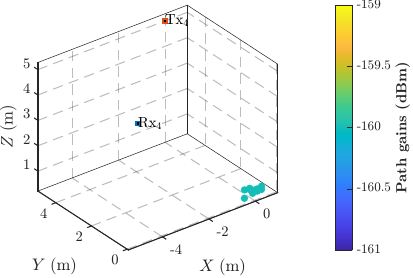}%
    \label{fig:mc1m4}%
  }

  \vspace{0.5em}

  \subfloat[View 5]{%
    \includegraphics[width=0.24\textwidth]{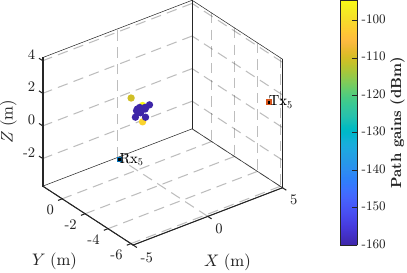}%
    \label{fig:mc1m5}%
  }\hfill
  \subfloat[View 6]{%
    \includegraphics[width=0.24\textwidth]{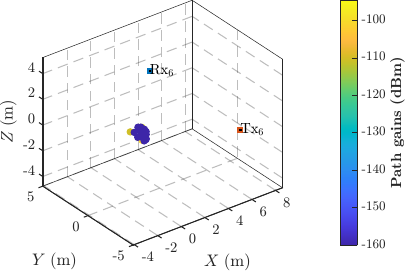}%
    \label{fig:mc1m6}%
  }\hfill
  \subfloat[View 7]{%
    \includegraphics[width=0.24\textwidth]{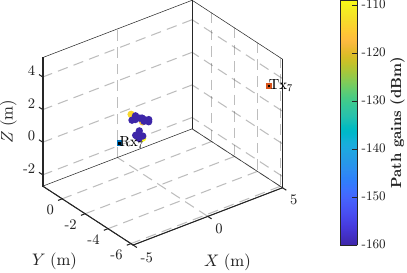}%
    \label{fig:mc1m7}%
  }\hfill
  \subfloat[\textcolor{black}{\ac{MVF}}]{%
    \includegraphics[width=0.24\textwidth]{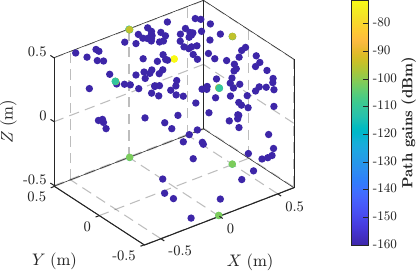}%
    \label{fig:mc1mfused}%
  }

  \caption{Eight representative views of the metal cube (1 m) scenario.}
  \label{fig:metal-cube-1m-collection}
\end{figure*}

\begin{figure*}[!t]
  \centering
  \subfloat[View 1]{%
    \includegraphics[width=0.24\textwidth]{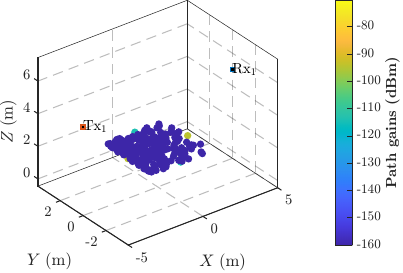}%
    \label{fig:mc4m1}%
  }\hfill
  \subfloat[View 2]{%
    \includegraphics[width=0.24\textwidth]{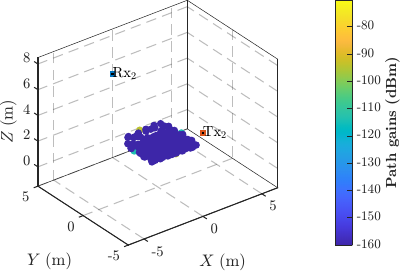}%
    \label{fig:mc4m2}%
  }\hfill
  \subfloat[View 3]{%
    \includegraphics[width=0.24\textwidth]{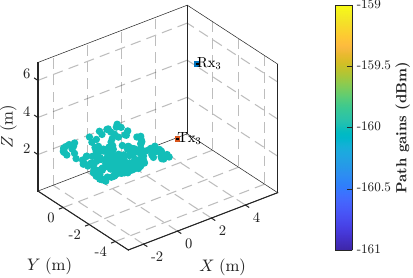}%
    \label{fig:mc4m3}%
  }\hfill
  \subfloat[View 4]{%
    \includegraphics[width=0.24\textwidth]{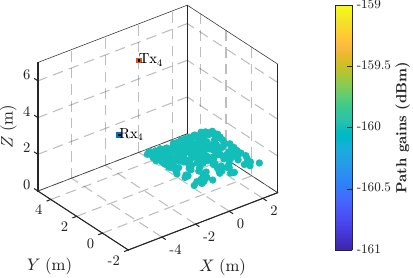}%
    \label{fig:mc4m4}%
  }

  \vspace{0.5em}

  \subfloat[View 5]{%
    \includegraphics[width=0.24\textwidth]{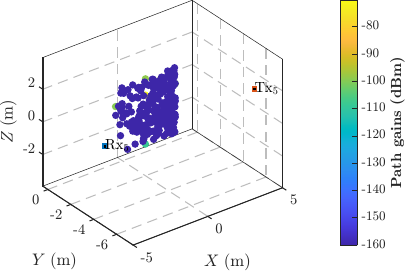}%
    \label{fig:mc4m5}%
  }\hfill
  \subfloat[View 6]{%
    \includegraphics[width=0.24\textwidth]{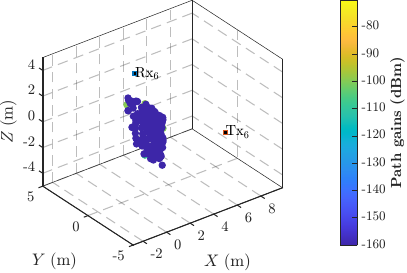}%
    \label{fig:mc4m6}%
  }\hfill
  \subfloat[View 7]{%
    \includegraphics[width=0.24\textwidth]{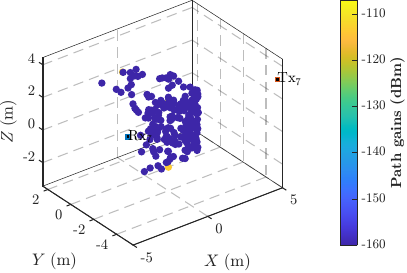}%
    \label{fig:mc4m7}%
  }\hfill
  \subfloat[\textcolor{black}{\ac{MVF}}]{%
    \includegraphics[width=0.24\textwidth]{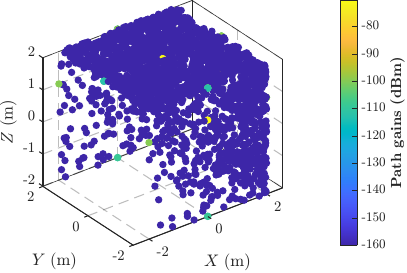}%
    \label{fig:mc4mfused}%
  }

  \caption{Eight representative views of the metal cube (4 m) scenario.}
  \label{fig:metal-cube-4m-collection}
\end{figure*}

Fig.~\ref{fig:metal-cube-1m-collection} presents $8$ reconstructions from different  TX-RX positions for the $3$D reconstructions of the $1$ m metal cube generated by the proposed imaging method at $6.75$ GHz.
Fig.~\ref{fig:mc1m1} throughout Fig.~\ref{fig:mc1m7} correspond to different TX-RX geometries.
Here, we notice how different TX-RX placements highlight complementary facets of the cube, whereby the brightest specular lobes consistently delineate the illuminated edges and corners while more modest returns appear along face in views, revealing diffraction and secondary scattering. When all seven outputs are fused in Fig.~\ref{fig:mc1mfused}, the result is an almost complete 3D cube that  delineates the cube's surface. 
The \textcolor{black}{\ac{MVF}} demonstrates the capability of the method that captures both the dominant specular lobes and the subtler scattering phenomena, yielding a high-fidelity reconstruction of metallic geometry from limited measurements, which aid in producing a metallic surface map from a sparse set of measurements.
In comparing the $1$ m metal cube of Fig.~\ref{fig:metal-cube-1m-collection} with the $4$ m cube of Fig.~\ref{fig:metal-cube-4m-collection}, we observe that the proposed imaging framework scales with object size.
For the larger cube, individual views in Fig.~\ref{fig:mc4m1} to Fig.~\ref{fig:mc4m7} still recover strong specular returns from each illuminated face, where the resulting point clouds are spread over a roughly four-times larger volume, reflecting the increased dimensions of the cube.
When the seven reconstructions are fused in Fig.~\ref{fig:mc4mfused}, they form a quasi-continuous $3$D cube approximately 4 m per side, in direct analogy to the near-closed shell of Fig.~\ref{fig:mc1mfused}.
Although larger objects demand proportionally wider angular coverage to illuminate every facet, the resolution of the method and fusion strategy consistently deliver near closed-surface reconstructions across scales.

\begin{figure*}[!t]
  \centering
  \subfloat[View 1]{%
    \includegraphics[width=0.24\textwidth]{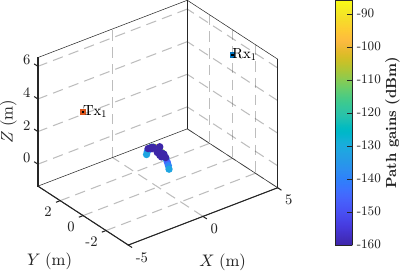}%
    \label{fig:mcp1}%
  }\hfill
  \subfloat[View 2]{%
    \includegraphics[width=0.24\textwidth]{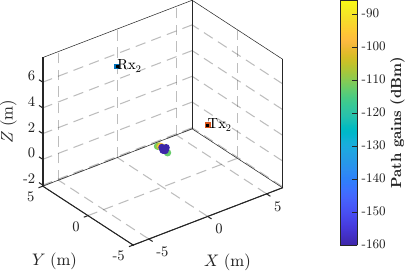}%
    \label{fig:mcp2}%
  }\hfill
  \subfloat[View 3]{%
    \includegraphics[width=0.24\textwidth]{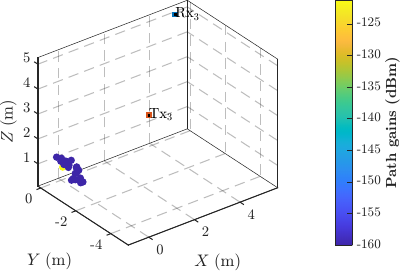}%
    \label{fig:mcp3}%
  }\hfill
  \subfloat[View 4]{%
    \includegraphics[width=0.24\textwidth]{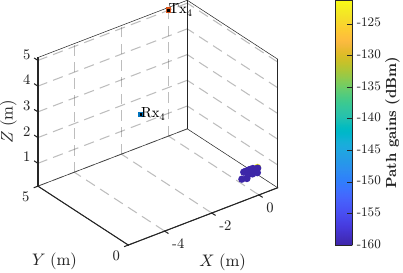}%
    \label{fig:mcp4}%
  }

  \vspace{0.5em}

  \subfloat[View 5]{%
    \includegraphics[width=0.24\textwidth]{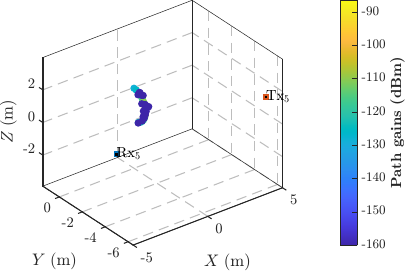}%
    \label{fig:mcp5}%
  }\hfill
  \subfloat[View 6]{%
    \includegraphics[width=0.24\textwidth]{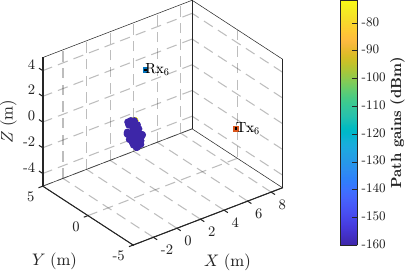}%
    \label{fig:mcp6}%
  }\hfill
  \subfloat[View 7]{%
    \includegraphics[width=0.24\textwidth]{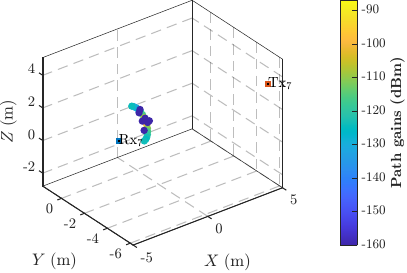}%
    \label{fig:mcp7}%
  }\hfill
  \subfloat[\textcolor{black}{\ac{MVF}}]{%
    \includegraphics[width=0.24\textwidth]{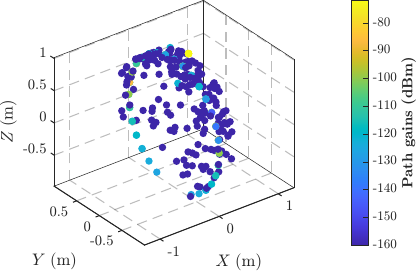}%
    \label{fig:mcpfused}%
  }

  \caption{Eight representative views of the metal circle plate scenario.}
  \label{fig:metal-circle-plate-collection}
\end{figure*}

Fig. \ref{fig:metal-circle-plate-collection} shows $3$D point clouds reconstructed by our imaging approach for the metal circle plate at $6.75$ GHz, with each point shaded by its recovered path gain in Fig. \ref{fig:mcp1} throughout Fig. \ref{fig:mcp7}, which correspond to seven different TX-RX configurations. 
In every view, the strongest echoes tightly hug the curved rim of the plate, which represents dominant specular reflection from the edge, while the weaker returns appear on the flat face or even the underside, depending on the illumination angle and propagation path. When the seven partial reconstructions are aggregated in Fig. \ref{fig:mcpfused}, they coalesce into an almost unbroken toroidal ring of scatterers, delineating both the circumference of the plate and its finite thickness which shows how fusion transforms incomplete single view snapshots into a detailed $3$D model of circular geometry from a sparse set of channel measurements.

\begin{figure*}[!t]
  \centering
  \subfloat[View 1]{%
    \includegraphics[width=0.24\textwidth]{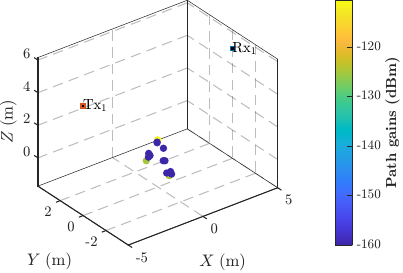}%
    \label{fig:mtp1}%
  }\hfill
  \subfloat[View 2]{%
    \includegraphics[width=0.24\textwidth]{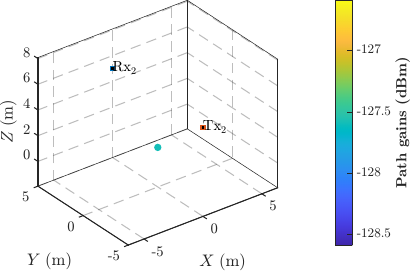}%
    \label{fig:mtp2}%
  }\hfill
  \subfloat[View 3]{%
    \includegraphics[width=0.24\textwidth]{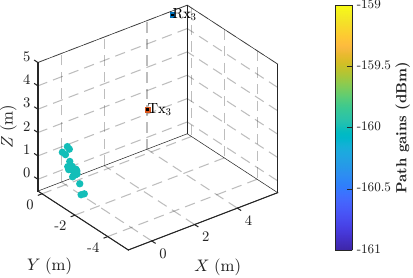}%
    \label{fig:mtp3}%
  }\hfill
  \subfloat[View 4]{%
    \includegraphics[width=0.24\textwidth]{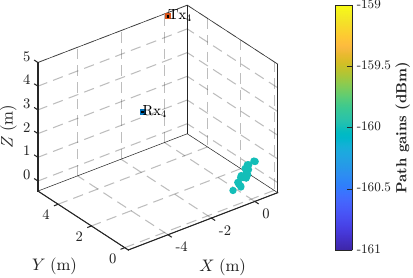}%
    \label{fig:mtp4}%
  }

  \vspace{0.5em}

  \subfloat[View 5]{%
    \includegraphics[width=0.24\textwidth]{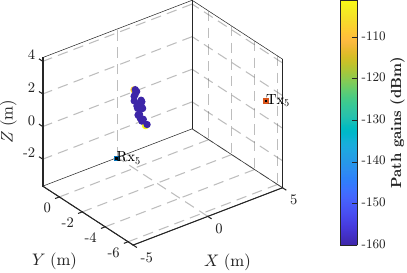}%
    \label{fig:mtp5}%
  }\hfill
  \subfloat[View 6]{%
    \includegraphics[width=0.24\textwidth]{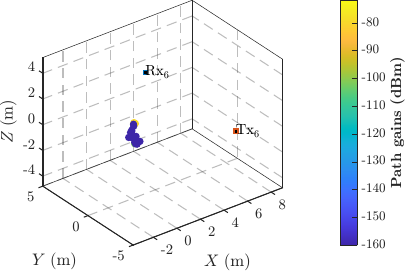}%
    \label{fig:mtp6}%
  }\hfill
  \subfloat[View 7]{%
    \includegraphics[width=0.24\textwidth]{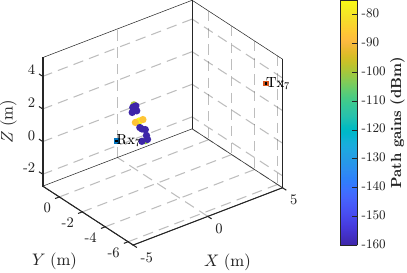}%
    \label{fig:mtp7}%
  }\hfill
  \subfloat[\textcolor{black}{\ac{MVF}}]{%
    \includegraphics[width=0.24\textwidth]{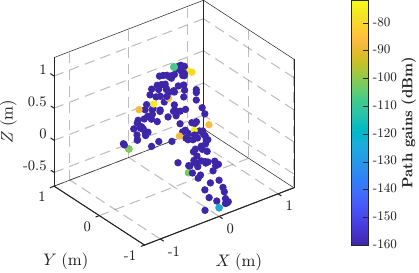}%
    \label{fig:mtp_fused}%
  }

  \caption{Eight representative views of the metal triangle plate scenario.}
  \label{fig:metal-triangle-plate-collection}
\end{figure*}

Fig. \ref{fig:metal-triangle-plate-collection} shows representative 3‑D reconstructions of the metal triangle plate scenario, generated by the proposed imaging algorithm and colored by the corresponding path gains.
Fig. \ref{fig:mtp1} throughout Fig. \ref{fig:mtp7} correspond to seven distinct TX-RX   positions.
We notice that the most intense returns consistently localize along two of the three linear edges of the plate, directly validating the capability of the algorithm to pinpoint dominant specular reflections from individual facets.
Simultaneously, the weaker scatterers that appear on the planar surface of the plate and around its apex, demonstrating its sensitivity to higher-order diffraction.

\begin{figure*}[!t]
  \centering
  \subfloat[View 1]{%
    \includegraphics[width=0.24\textwidth]{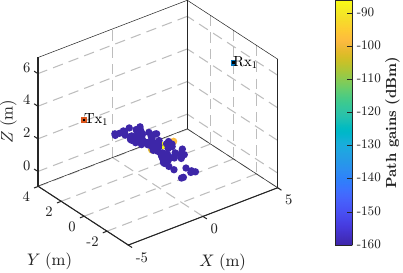}%
    \label{fig:tesla1}%
  }\hfill
  \subfloat[View 2]{%
    \includegraphics[width=0.24\textwidth]{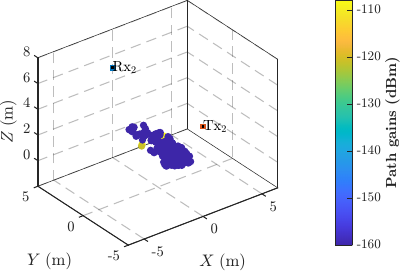}%
    \label{fig:tesla2}%
  }\hfill
  \subfloat[View 3]{%
    \includegraphics[width=0.24\textwidth]{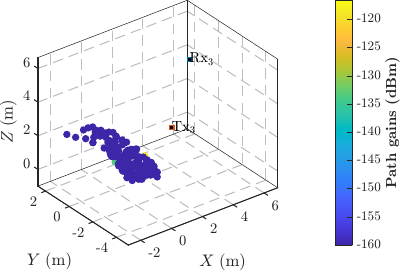}%
    \label{fig:tesla3}%
  }\hfill
  \subfloat[View 4]{%
    \includegraphics[width=0.24\textwidth]{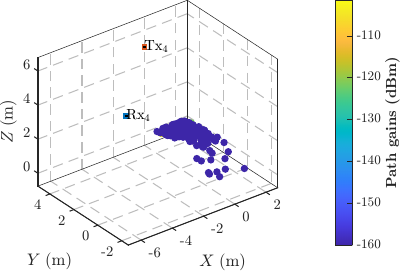}%
    \label{fig:tesla4}%
  }

  \vspace{0.5em}

  \subfloat[View 5]{%
    \includegraphics[width=0.24\textwidth]{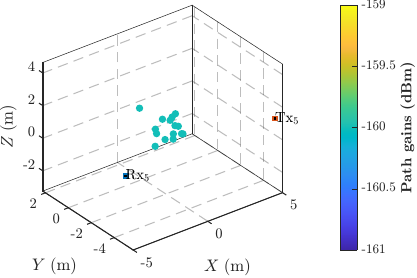}%
    \label{fig:tesla5}%
  }\hfill
  \subfloat[View 6]{%
    \includegraphics[width=0.24\textwidth]{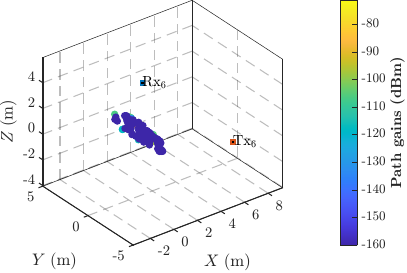}%
    \label{fig:tesla6}%
  }\hfill
  \subfloat[View 7]{%
    \includegraphics[width=0.24\textwidth]{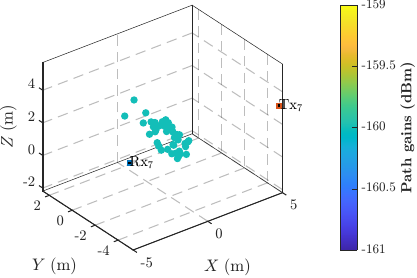}%
    \label{fig:tesla7}%
  }\hfill
  \subfloat[\textcolor{black}{\ac{MVF}}]{%
    \includegraphics[width=0.24\textwidth]{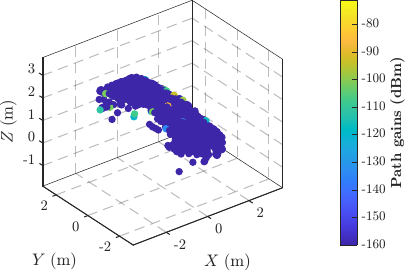}%
    \label{fig:tesla_fused}%
  }

  \caption{Eight representative views of the Tesla scenario.}
  \label{fig:tesla-collection}
\end{figure*}

Fig. \ref{fig:tesla-collection} showcases various $3$D reconstructions of the Tesla  vehicle model.
In particular, Fig.~\ref{fig:tesla1} to Fig.~\ref{fig:tesla7} correspond to seven unique TX-RX positions. 
In Fig.~\ref{fig:tesla1}, the brightest returns concentrate on the hood and windshield, highlighting strong specular reflections from the smooth surfaces.  
In Fig.~\ref{fig:tesla3} and Fig.~\ref{fig:tesla4}, the view shifts focus toward the roofline, where high gain lobes delineate the sides of the vehicle.
Meanwhile, Fig. \ref{fig:tesla5} to Fig. \ref{fig:tesla7} reveal secondary, lower amplitude echoes along the side panels, and wheel arches.
Upon fusing all seven point clouds in Fig. \ref{fig:tesla_fused}, the result is a richly detailed, continuous $3$D body of the Tesla car silhouete and roof, demonstrating how the fusion strategy can reconstruct automotive shapes from a handful of limited view measurements.

\begin{figure*}[!t]
  \centering
  \subfloat[Standard tree model]{%
    \includegraphics[width=0.32\textwidth]{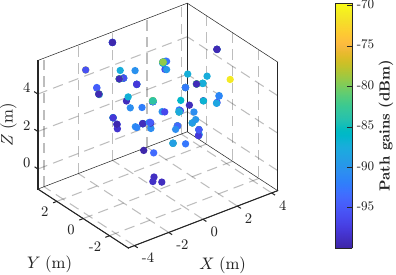}%
    \label{fig:dynamic-range1}%
  }\hfill
  \subfloat[Metal Cube (\textcolor{black}{1 m}) model]{%
    \includegraphics[width=0.32\textwidth]{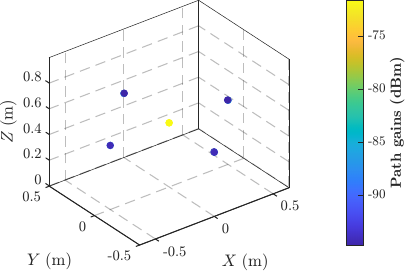}%
    \label{fig:dynamic-range2}%
  }\hfill
  \subfloat[Metal Cube (\textcolor{black}{4 m}) model]{%
    \includegraphics[width=0.32\textwidth]{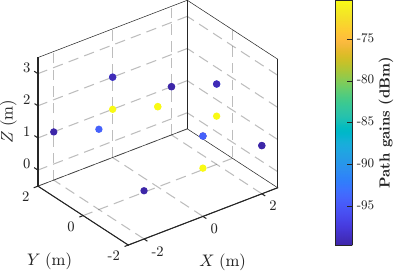}%
    \label{fig:dynamic-range3}%
  }\hfill
  \subfloat[Metal Triangle model]{%
    \includegraphics[width=0.32\textwidth]{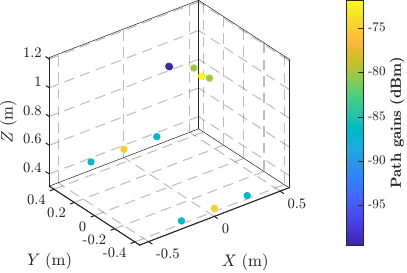}%
    \label{fig:dynamic-range4}%
  }\hfill
  \subfloat[Metal Circle model]{%
    \includegraphics[width=0.32\textwidth]{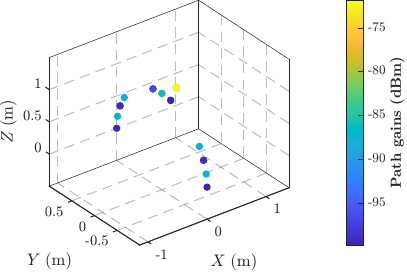}%
    \label{fig:dynamic-range5}%
  }\hfill
  \subfloat[Tesla model]{%
    \includegraphics[width=0.32\textwidth]{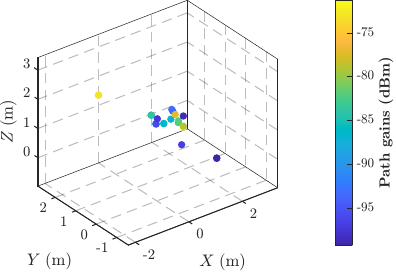}%
    \label{fig:dynamic-range6}%
  }\hfill
  \caption{The reconstructed images after introducing noise floor of $-100$ dBm for limited dynamic range.}
  \label{fig:dynamic-range}
\end{figure*}

To have realistic dynamic range on the ray tracer, in Fig. \ref{fig:dynamic-range} we introduce a noise floor of $-100$ dBm, which means that components weaker than \(-100\,\text{dBm}\) are suppressed.
The geometry of the six test objects in Fig.~\ref{fig:dynamic-range1} throughout  Fig.~\ref{fig:dynamic-range5} is preserved, depending on the object.
The foliage canopy and the circular arc retain their spatial outline, whereas the car model, metal cubes and the metal triangle need additional TX-RX views.

\begin{figure}[!t]              
  \centering
  \includegraphics[width=0.4\textwidth]{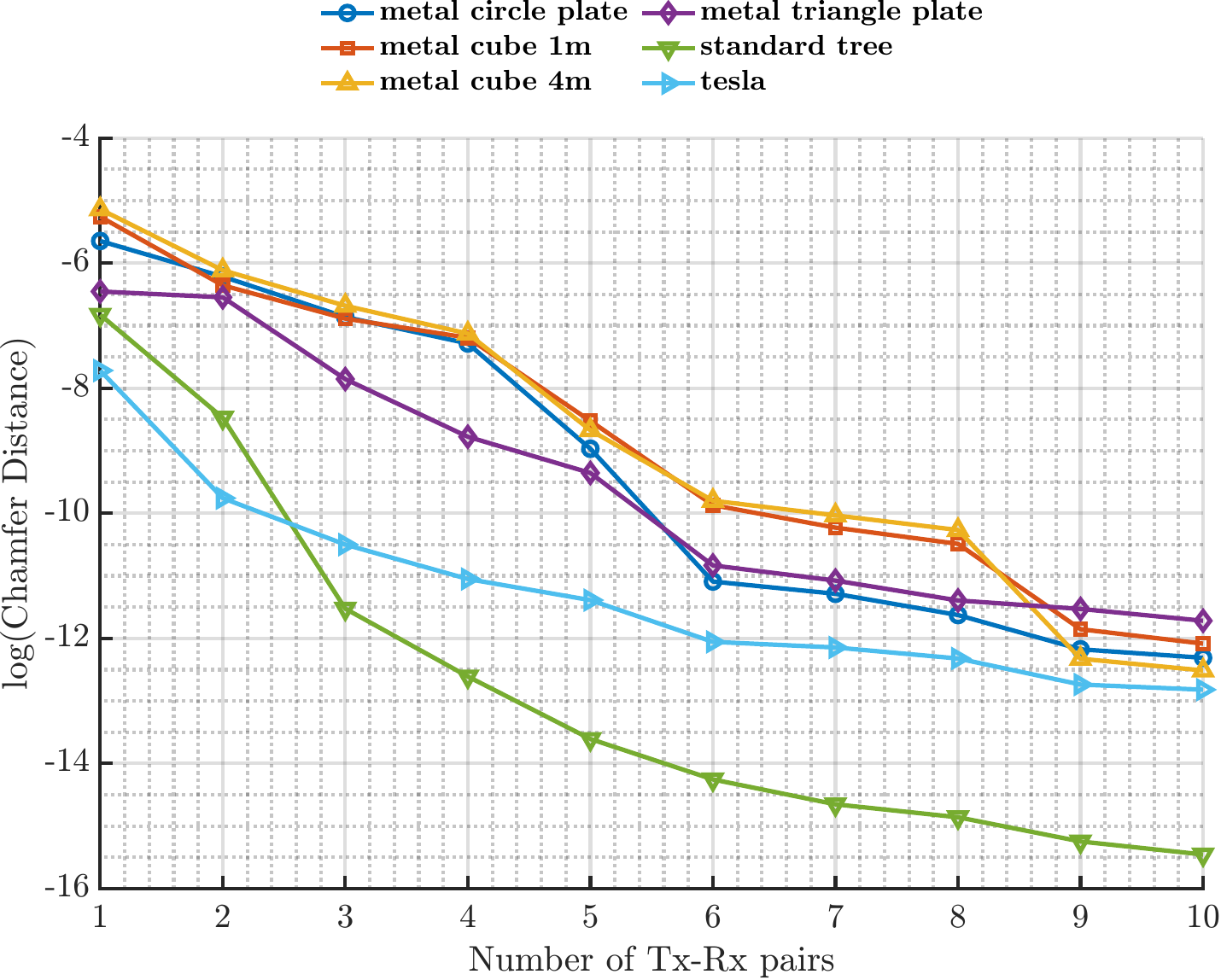}
  \caption{\textcolor{black}{Tradeoffs between Chamfer distance and number of Tx-Rx pairs for various objects.}}
  \label{fig:chamfer}
\end{figure}

\textcolor{black}{To study the tradeoffs between imaging accuracy and the required number of TX-RX pairs, it is crucial to select an appropriate metric for imaging accuracy. 
\ac{CD} is a universal metric to measure of dissimilarity between point clouds \cite{wu2021balanced} and based on nearest neighboring points, which is given as 
\begin{equation}
	d_{C D}=\frac{1}{\left|\mathcal{S}_\gamma\right|} \sum_{\mathbf{x} \in \mathcal{S}_\gamma} \min _{\mathbf{y} \in \mathcal{S}}\|\mathbf{x}-\mathbf{y}\|_2+\frac{1}{\left|\mathcal{S}\right|} \sum_{\mathbf{y} \in \mathcal{S}} \min _{\mathbf{x} \in \mathcal{S}_\gamma}\|\mathbf{y}-\mathbf{x}\|_2,
\end{equation}
where $\mathcal{S}_\gamma$ and $\mathcal{S}$ are the reconstructed one conditioned on $\gamma$ and the reference cloud accurately describing the object image, respectively.
In Fig. \ref{fig:chamfer}, we plot the \ac{CD} (on $\log_2$ scale) vs the number of Tx-Rx pairs used for the \ac{MVF}, where a clear decaying trend in \ac{CD} is shown with increasing number of Tx-Rx pairs, which favors a good tradeoff between number of Tx-Rx pairs and \ac{CD} for proper reconstruction.  
For flat metallic surfaces, such as metal circle plate and metal triangle plate, we observe that after roughly $6$ Tx-Rx pairs, the \ac{CD} drastically drops to about $2^{-11}$ and adding more Tx-Rx pairs yields minor improvements of about $2^{0.3}$ on linear scale for each additional Tx-Rx pair.
The trends of metallic cube structure are similar irrespective of its dimension. 
On the other hand, complex objects such as the Tesla vehicule and the standard tree exhibit different trends as compared to flat and cubic structures. In particular, we observe that after $3$ Tx-Rx pairs, the \ac{CD} decays with a steady slope then converges towards an almost straight line for the standard tree, whereas it plateaus for the Tesla, indicating that $8$ to $10$ Tx-Rx pairs may be enough for a good reconstruction. It is also worth noting that the standard tree exhibits the greatest reduction of about $2^{8.6}$ on a linear scale.

}

\section{Open Challenges}
\label{sec:open-challenges}
While the \textit{\textcolor{black}{\ac{ERP}}} approach demonstrates the capability of imaging objects through \ac{CSI} information via NYURay ray tracing experiments, several important challenges remain to bring \ac{ISAC} imaging into real-world practice:
\begin{itemize}
\item  \textbf{Real-time computational efficiency:} Ray tracing and reflection point optimization demand substantial computational resources, presenting challenges for ISAC deployment. Future research must explore simplified diffraction models as alternatives to computationally intensive Maliuzhinets-based approaches \cite{wang2005parametric}. Efficient approximations preserving essential electromagnetic characteristics while reducing processing requirements could enable RF imaging in 6G systems with sub-millisecond latency constraints \cite{10901856, 10494224}. Hardware advancements following Moore's law combined with optimized propagation algorithms will enable wireless devices to perform site-specific channel prediction for dynamic beam management in vehicular communications and XR applications requiring ultra-reliable low-latency connections \cite{10489861, 10726912}.
\textcolor{black}{It is worth mentioning that in contrast to purely image based ray tracers, NYURay's hybrid shooting bouncing rays combined with image-based ray tracing reduces the computational overhead in comparison to image-based ray tracing \cite{10721240}. In addition, the computational requirements of shooting bouncing ray methods scale linearly with the number of obstructions in a given environment \cite{600436}. Recent techniques have also been proposed based o	n the shooting bounce methods, such as the vertical-plane-launch technique introduced in \cite{686774} for site-specific ray tracing in order to predict propagation effects for the $300$ $\operatorname{MHz}$–$3$ $\operatorname{GHz}$ band. Therefore, future $6$G \ac{ISAC} frameworks may build on top of shooting bounce rays to develop faster ray tracing methods that are deemed computationally fast for usage.}
\item \textbf{Synthetic Aperture Radar inspired \ac{RF} images.} 
\textcolor{black}{Inspired by the way a \ac{SAR} satellite leverages its own motion to create an enormous effective aperture, we ask the parallel question: \emph{Can channel-state information (\ac{CSI}) gathered across time and distance be fused to synthesize a much broader effective bandwidth, and thus deliver higher-resolution RF imagery?} The \ac{FR3}, with its inherently multi-band spectrum, offers just the spectral headroom needed for such a virtual bandwidth expansion \cite{bazzi2025upper}.}
\item \textbf{On bandwidth and \ac{SNR}.} The proposed imaging technique assumed perfect \ac{CSI} information at for each TX-RX position pair. In practice, the transmissions are limited by bandwidth and \ac{SNR}, which can further limit the \ac{CSI} estimation accuracy, \textcolor{black}{followed by the sensing parameters per path component.} Robust estimators making use of possible moving apertures or/and integration times, may be used to further enhance the accuracy of \ac{CSI} estimates. Therefore, a natural question to ask is \textit{how much bandwidth and \ac{SNR} does one require to generate a high-resolution \ac{RF} image?} Even more, \textit{is there any one-to-one mapping between the bandwidth and the \ac{RF} image resolution one can achieve?}
\textcolor{black}{In a practical \ac{OFDM} scenario, the bandwidth is one fundamental requirement for imaging because it dictates the finest resolvable path delay which is inversely proportional to the bandwidth, allowing specular components that were previously inseparable to appear as distinct taps in the estimated channel impulse response.
Synthetic bandwidth range profiling method for moving targets has been proposed in \cite{li2021synthetic} to achieve extra bandwidth.
In addition, the multiband spectrum given on the upper midband FR3 can also favor  resolvability of similar path structures in the delay domain. 
Besides the temporal domain, the number of transmitting and receiving antennas required to spatially separate similar path structures can also play a crucial role in providing high separation of the different paths. 
For example, when two path components reach the receiver at almost the same delay but from different directions, the spatial diversity of the transmit- and receive-antenna arrays can allow path distinguishability.
However, the question of how much bandwidth, antennas and movement is needed to form a clear image of an intended object, and how the aforementioned resources relate to imaging metrics, such as the \ac{CD} and the \ac{SSIM} remain an open question for further investigation.}
\item \textcolor{black}{\textbf{Aggregate cost for imaging.}} Our \textit{\textcolor{black}{\ac{ERP}}} estimate successfully recovers \textcolor{black}{two bounce returns} on a path-by-path basis. It can be interesting to form an aggregate cost $\sum_i \sum_j\left|g_{i, j}\right|^2 f\left(\alpha^{(i, j)}, \beta^{(i, j)}\right)$ to take into account path reliabilities when forming an \ac{RF} image, so that weaker, low-gain paths contribute proportionally less to the final \ac{RF} image.
\textcolor{black}{It should be noted that the simulation of rays is not required for the reconstruction algorithm to function. 
Rays generated by a ray tracer provide a highly realistic means of modeling an object.}
\item \textbf{Model mismatch and hardware impairments:} Phase noise, mutual coupling, and calibration errors in real phased arrays distort the measured \ac{CSI}. Robust imaging must incorporate models of the imperfections or jointly estimate the hardware impairments alongside object geometries.
 \item \textbf{Integration with digital twin workflows:} Finally, coupling our imaging algorithm with on-the-fly digital twin and multi-modal frameworks in addition to learned scene priors, e.g. from photogrammetric maps, can improve \ac{RF} images in complex urban or indoor settings.
 \item \textcolor{black}{\textbf{Outlier detection of environmental points:}
Some objects are not total reflectors, such as humans, and in some cases, a path can completely penetrate the object hence providing little-to-no information about the object in hand. 
Our current framework can be used in conjunction with an outlier detection mechanism that processes all \acp{ERP}, in addition with prior information about the indoor environment. In particular, given the geometry of the indoor environment, one can compute the set of \acp{ToA}, \acp{AoA}, \acp{ZoA}, \acp{AoD}, \acp{ZoD} that corresponds to the map of the environment. The environmental point cloud will include the paths that directly pass through the objects. With the help of the points corresponding to the environment, the set of point cloud $\mathcal{S}_\gamma$ representing the object in question can be post-processed using an outlier detection mechanism to clean all \acp{ERP} that are close to environmental point clouds. Therefore, future work should also propose sophisticated outlier detection mechanisms to reject points that cluster around the environmental points.}
\end{itemize}

\section{\textcolor{black}{Conclusion \& Future Work}}
\label{sec:conclusion}
In this paper, we presented a novel end-to-end \ac{ISAC} imaging pipeline for the $6.75$ GHz mid-band. We generate \ac{CSI} per-path components by the site-specific NYURay ray tracing engine and feed it to a physics-guided inverse solver that localizes equivalent reflection points representative of objects of different shapes and size.
Central to our approach is an optimization framework that jointly estimates transmit and receive segments via path delays and angles, from which we infer equivalent reflection points capturing a single-point representation of multi-bounce paths.
Our framework also performs an \textcolor{black}{\ac{MVF}} step from different TX-RX positions in order to circumvent possible limited angular coverage to yield dense $3$D point clouds for diverse object types, including $1$ m and $4$ m metal cubes, vehicles, circular and triangular plates, and standard trees.
Our results demonstrate how the proposed method scales with object size,  reconstructs planar faces, edges and corners, all without assuming known surfaces given \ac{CSI} information.
\textcolor{black}{Although the six-tuple parameters were generated with NYURay, we demonstrate that identical parameters can be recovered from standard-compliant CSI-RS/SRS measurements defined in 3GPP TS 38.211/214 (5G NR) and TS 36.211/214 (LTE) via high-resolution estimators; consequently, the proposed RF-imaging pipeline is then applicable.
Looking ahead, the ray-tracing driven imaging framework paves the way for real-time dynamic object imaging and fusion on \ac{FR3} and establishes a practical path toward \emph{lens-free} $6$G sensing solutions that coexist seamlessly with high-capacity communications for future \ac{ISAC} systems.}
\textcolor{black}{For future work, we will consider benchmarking our NYURay performance against full-wave electromagnetic simulation methods.
While our current ray tracer does not explicitly model multiple scattering within dielectric objects, such effects typically result in significantly attenuated multipath components after multiple penetrations and internal reflections. These weak higher-order contributions, which can be captured by conventional full-wave scattering algorithms, such as inverse scattering methods \cite{chen2018computational,dubey2022phaseless}, are expected to have limited impact on our sensing results, as the ISAC system primarily relies on stronger direct-path and specular reflection components.}

\bibliographystyle{IEEEtran}
\bibliography{refs}

\vfill

\end{document}